\documentclass[preprint,12pt]{elsarticle}

\usepackage{amssymb}
\usepackage{amsmath}
\usepackage[utf8]{inputenc}
\usepackage[english]{babel}
\usepackage{graphicx}
\usepackage{sidecap} 
\usepackage[unicode]{hyperref}
\usepackage[tableposition=top]{caption}
\usepackage{multirow}
\usepackage{textcomp}
\usepackage{gensymb}
\usepackage{graphics}
\usepackage{color}
\usepackage[dvipsnames]{xcolor}
\usepackage[margin=1in]{geometry}
\usepackage{chngcntr}
\usepackage{soul}
\usepackage{fancyhdr}
\usepackage{tikz}
\usepackage{orcidlink}

\journal{Vacuum}

\begin{document}

\pagestyle{fancy}
\fancyhead{} 
\fancyhead[L]{\textbf{\tiny This paper has been accepted for publication in Vacuum. This is the author's version which has not been fully edited and content may be different from the final publication. Citation information: F. do Nascimento, Ananias A. Barbosa, and K. G. Kostov, Vacuum, vol. 246, p. 115027, Mar. 2026, DOI \href{https://dx.doi.org/10.1016/j.vacuum.2025.115027}{10.1016/j.vacuum.2025.115027}}}

\begin{frontmatter}

\title{Water activation using $\rm{Ar-H_2}$ atmospheric pressure plasma jets}


\author{Fellype do Nascimento\corref{cor1} \orcidlink{0000-0002-8641-9894}}
\author{Ananias Alves Barbosa \orcidlink{0000-0001-5588-5844}}
\author{Konstantin Georgiev Kostov \orcidlink{0000-0002-9821-8088}}

\cortext[cor1]{corresponding author: fellype@gmail.com}
\affiliation{organization={Faculty of Engineering and Sciences in Guaratinguetá, São Paulo State University–UNESP},
            addressline={}, 
            city={Guaratinguetá},
            postcode={12.516-410}, 
            state={São Paulo},
            country={Brazil}}

%

\begin{abstract}
Whether for materials processing or medical applications, the use of atmospheric pressure plasma jets (APPJs) has emerged as a relevant alternative to conventional methods. Within the APPJs research field, the search for innovation aims not only to solve existing problems but also to explore novel options for generating plasma jets and find new possible applications. In this work, the properties of $\rm{Ar-H_2}$ APPJs generated using two plasma sources that differ in the frequency, amplitude, and waveform of the generated voltage signal were studied through electrical, thermal, and optical characterization. The discharge parameters were analyzed as a function of the $\rm{H_2}$ content in the gas mixture, with this parameter varying from 0\% to 3.5\%. Optical emission spectroscopy revealed that the same reactive species were produced for both plasma sources, except nitric oxide ($\rm{NO}$), which was observed only for the source operated at a higher frequency (PS \#1). Applications for water activation were performed without $\rm{H_2}$ and with 3.5\% $\rm{H_2}$ in the gas mixture. The results of water treatment revealed that ammonia is also produced when $\rm{H_2}$ is added to the working gas. This finding suggests that the water treated by a $\rm{Ar-H_2}$ plasma jet can be an attractive option for use in agriculture.
\end{abstract}


\end{frontmatter}

\begin{tikzpicture}

\node[align=justify, text width=0.9*\textwidth, inner sep=1em]{
{\small This paper has been accepted for publication in Vacuum. This is the author's version which has not been fully edited and content may be different from the final publication. Citation information: F. do Nascimento, Ananias A. Barbosa, and K. G. Kostov, Vacuum, vol. 246, p. 115027, Mar. 2026, DOI \href{https://dx.doi.org/10.1016/j.vacuum.2025.115027}{10.1016/j.vacuum.2025.115027}}
};

\node[xshift=3ex, yshift=-0.7ex, overlay, fill=white, draw=white, above
right] at (current bounding box.north west) {
\textit{Dear reader,}
};

\end{tikzpicture}

\section{Introduction}

Research involving atmospheric pressure plasmas (APPs) has received increasing attention in recent years. The advances achieved in this technology have enabled its utilization in diverse areas, spanning from material treatment and food processing to biological applications such as in medicine and dentistry \cite{Bartis_interaction_2016, Deepak_review_2022, Adesina_review_2024, Jeong_clinical_2024, koga-ito_cold_2024, Weihe_microwave_2024}. {In particular, atmospheric pressure plasma jets (APPJs) have the ability to operate directly in the clinical or environmental setting, without the need of a vacuum infrastructure, as is the case in most industrial plasma applications.} One of the main features that allows the usage of APPs in biology is theow gas temperature of this category of plasmas, usually around room temperature \cite{gerling_relevant_2018}. However, the success of applications of APPs in medicine and dentistry is attributed to the interaction of the reactive oxygen and nitrogen species (RONS) produced by the plasma with the biological targets \cite{Miebach_medical_2022, Ghasemitarei_effects_2023, koga-ito_cold_2024}. One of the most common ways of producing RONS is to generate a plasma jet using a noble gas in a way that it interacts with the ambient air \cite{Montalbetti_production_2025, Bende_reactive_2025}. The kind of reactive species produced by an APP depends on factorsike working gas composition, voltage waveform/amplitude/frequency, device configuration, target characteristics such as conductivity, humidity, material, and so on \cite{Sobota_diagnostic_2024}. Nevertheless, when operating in open environments, the most commonly reported RONS are hydroxyl ($\rm{OH}$), nitric oxide ($\rm{NO}$), molecular nitrogen ions ($\rm{N_2^{+}}$), and oxygen atoms ($\rm{O}$) \cite{Woedtke_foundations_2022}. Other reactive speciesike nitrogen atoms ($\rm{N}$), cyanide ($\rm{CN}$), and carbon monoxide ($\rm{CO}$) are also reported in theiterature \cite{Ridenti_cn_2018, Mestre_comparison_2024}. All those RONS are easily observed using optical emission spectroscopy (OES) of the plasma/plasma jet. However, APPs can produce many more RONS such as hydrogen peroxide ($\rm{H_2O_2}$), nitrite ($\rm{NO_2}$), nitrate ($\rm{NO_3}$), and ozone, which can only be detected with specific instrumentsike gas analyzers, optical absorption spectroscopy (OAS), and Fourier transform infrared (FTIR) spectroscopy. Nitrous acid ($\rm{HNO_2}$), nitric acid ($\rm{HNO_3}$), nitrous oxide ($\rm{N_2O}$), and dinitrogen pentoxide ($\rm{N_2O_5}$) are also reported in some works \cite{reuter_kinpenreview_2018, Zhou_dynamics_2024}.

In many cases, to enhance the production of RONS in APPs, different gas mixtures have been employed. Most commonly, they consist of a noble gas with the addition of small amounts of oxygen, nitrogen, air, or water vapor \cite{Montalbetti_production_2025, Bende_reactive_2025, Bradu_reactive_2020}. 
Previous studies on plasma treatment of polymers, water, and biological tissues \cite{Bende_reactive_2025, Armenise_atmospheric_2020, Puac_plasmaliquid_2025} reported that the use of gas mixtures instead of pure gases hased to improved results.  For those specific purposes, the $\rm{Ar-H_2}$  gas mixture has not been sufficiently explored.

{The addition of $\rm{H_2}$ to $\rm{Ar}$ plasmas produces significant modifications in both discharge characteristics and application performance. Such modifications are notably different when plasmas are produced atow pressure or at atmospheric pressure.} \cite{Wu_effects_2013, Synek_interplay_2015, ellis_effects_2016, Guo_characteristics_2018, mauer_how_2021, Cho_experimental_2025}
{Forow‑pressure plasma, even small $\rm{H_2}$ fractions considerably alter the plasma behavior. Hydrogen addition changes jetength, electron temperature, plasma density, plasma chemistry, enthalpy, thermal conductivity, and recombination rates.} \cite{ellis_effects_2016, mauer_how_2021, Cho_experimental_2025}
{For APPs, the results show a relatively strong and non-monotonic dependence of the gas temperature ($T_g$) and electron density ($n_e$) on gas composition.} \cite{Synek_interplay_2015, Guo_characteristics_2018}
{Regarding the applications employing the $\rm{Ar-H_2}$, the presence of $\rm{H_2}$ in the gas mixture typicallyeads to an improvement in the results because of the chemical activity of the atomic hydrogen. For example, Nakahiro \textit{et al} reported better results of $\rm{Cu}$ film deposition using a $\rm{Ar-H_2}$ gas mixture in a proportion of 1000$-$10 sccm and Udachin \textit{et al} reported deoxidation and shielding effects on the surface exposed to the $\rm{Ar-H_2}$ plasma jet} \cite{Nakahiro_effect_2012, Udachin_shielding_2025}.

Some works employing APPJs with $\rm{H_2}$ added working gas reported the production of imidogen ($\rm{NH}$). Toda \textit{et al} employed a $\rm{N_2}-\rm{H_2}$ gas mixture in a pulsed-arc plasma jet. In that work, they measured the intensity of theight emitted by $\rm{NH}$ molecules as a function of the $\rm{H_2}$ fraction and observed a peak emission with $\sim$0.25\% of $\rm{H_2}$ in the mixture, followed by an exponential decrease as more $\rm{H_2}$ was added \cite{Toda_bright_2020}. $\rm{NH}$ radical was observed by Hamdan \textit{et al} using a $\rm{Ar-N_2}$ mixture in a microwave plasma jet submerged in water \cite{Hamdan_microwave_2018}. The $\rm{NH}$ radical was also reported inaser-induced plasmas using $\rm{N_2}$ and ambient air \cite{Choi_compensating_2024}. Fujera \textit{et al} \cite{Fujera_aerosol-based_2024} observed the production of $\rm{NH}$ using a multihollow surface DBD with the addition of water vapor to the working gas. On the other hand, other works on APPs with $\rm{Ar-H_2}$ did not report the presence of $\rm{NH}$ \cite{Guo_characteristics_2018, das_optical_2018}

In this work, a study of APPJs generated using a $\rm{Ar-H_2}$ mixture is presented. Basic properties of the $\rm{Ar-H_2}$ plasma jet were analyzed in comparison with the APPJ produced without $\rm{H_2}$ in the working gas. For this purpose, we employed two plasma sources with a similar geometrical configuration (mounting/structure), but driven by different power supplies. One of them is an AC voltage generator operating at a frequency of about 588 Hz, while the other one is a pulsed source with an effective pulse repetition rate of 180 Hz. The waveforms of each power supply also differ from each other. {The two plasma sources are intended for different applications: while the AC-driven jet is more appropriate for material surface modification, the pulsed plasma jet is more suitable for biomedical purposes. The utilization of two distinct plasma sources provided the opportunity to investigate which plasma properties are common for both devices and which characteristics depend on the specific excitation method. Furthermore, the introduction of $\rm{H_2}$ served as a key variable for the prospective adaptation of the plasma's chemical output for its application in water activation. The results indicate that electron density and vibrational temperature exhibit comparable behavior across both plasma sources, demonstrating source‑independent characteristics. In contrast, gas and rotational temperatures, as well as the total discharge power, display source‑dependent variations. These findings elucidate the factors governing species formation.}
For both devices, an increment in the gas temperature was observed as more $\rm{H_2}$ was added to the plasma. However, this effect is more pronounced when a power supply with a higher frequency is employed. In all the cases, when $\rm{H_2}$ is added to the working gas, the production of $\rm{NH}$ within the plasma jet is detected. Applications of $\rm{Ar}$ or $\rm{Ar-H_2}$ APPJs in water treatmentead to different results due to the distinct APPJ properties obtained for each gas composition. {It is important to note that the plasma treatment used in this work consists of applying APPJs directly to the surface of the water in an open environment. Therefore, the findings in this work areimited to ambient atmospheric processing conditions, and effects of vacuum or sub-atmospheric pressure were not investigated.}

\section{Materials and Methods}

\subsection{The plasma sources and setup overview}
\begin{figure}[htb]
\centering
\includegraphics[width=0.9\textwidth]{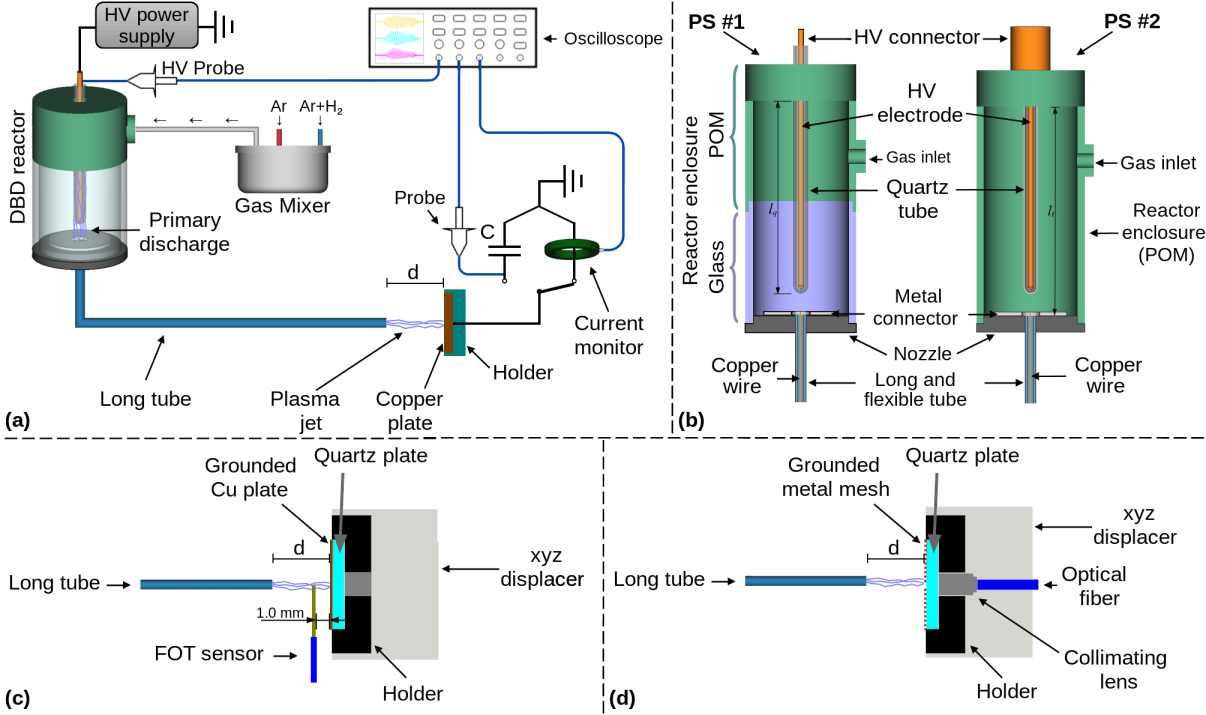}
\caption{Experimental arrangement. (a) Overview and electrical measurement scheme. (b) Reactor configuration for plasma sources \#1 and \#2,eft and right, respectively. (c) Setup for gas temperature measurements. (d) OES measurement scheme. The elements in the figures are out of scale. The value of $d$ was 10.0 mm in all experiments carried out in this work.\label{expSetup}}
\end{figure}

Figure~\ref{expSetup} shows the experimental arrangements used in this work. The Figure~\ref{expSetup}(a) gives an overview of the entire experimental setup for plasma jet generation and electrical data acquisition. Figure~\ref{expSetup}(b) presents detailed views of the dielectric barrier discharge (DBD) reactor, which integrates the plasma sources. Figures~\ref{expSetup}(c-d) show the schemes for measurements of gas temperature and optical emission spectroscopy. Experiments were carried out with two different plasma sources (PS \#1 and PS \#2 in Figure~\ref{expSetup}(b)) that have the same operation principle but different power supplies and reactor dimensions. Both plasma sources consist of a power supply and a DBD reactor to which a 1.0 m-long (with outer and inner diameters of 4.0 mm and 2.0 mm, respectively) flexible plastic tube (made ofow-density polyethylene) is connected. Along theong tube passes a thin copper wire (0.5 mm in diameter), which terminates 2 mm before the plastic tube exit. The other end of the wire is fixed to a metallic connector placed inside the reactor chamber. {The copper wire within the plastic tube, acting as a floating electrode, acquires the plasma’s electric potential, thus enabling the ignition of a small plasma jet at the far end of theong tube }\cite{Kostov_study_2015, do_nascimento_comparison_2017}. The DBD reactor includes a pin electrode made of tungsten encapsulated by a closed-end quartz tube. The dimensions of the DBD reactors and their inner elements are given in Table~\ref{reactorDim}. {The dimensions of the PS \#2 reactor enclosure were selected to ensure that the power value per unit volume of gas (considering only the useful internal volume) is approximately equivalent to that achieved in PS \#1.} Both devices employ the jet transfer technique to produce plasma jets at the tip of theong tube. The operating principle of this kind of device is described in previous works of our research groups \cite{Kostov_study_2015, do_nascimento_comparison_2017, do_Nascimento_different_2023}.

The plasma source \#1 was driven by a commercial power supply (Minipuls4 GBS Elektronik GmbH, Germany), connected to the pin electrode by a crocodile clamp. In this case, the applied voltage waveform is an amplitude modulated sinusoidal signal with an oscillation frequency ($f_{osc}$) of 27.0 kHz {($t_{on} \simeq 0.57$ ms)}, followed by a voltage off period {($t_{off} \simeq 1.13$ ms). The entire voltage burst} repeats at a repetition period ($\tau$) of 1.7 ms (frequency of nearly 588 Hz). The plasma source \#2 uses a portable power supply that was adapted from a device used in aesthetic applications (from Ibramed Ltda, Brazil). It generates damped-sine waveforms with peak voltage values that can reach up to 20 kV and an oscillation frequency of $\sim$110 kHz in the sinusoidal phase, {with effective duration (voltage higher than 5 kV) of nearly 70 {\textmu}s}. It generates an adjustable sequence of dumped high-voltage (HV) oscillations (3 in this work){ with a separation of $\approx$1.7 ms between each two pulses, being that the sequence of pulses is} repeated with theine frequency (60 Hz), producing an effective pulse repetition rate of 180 Hz. In this case, the pin-electrode is connected to a male metallic socket, which is attached to the dielectric enclosure and plugged into the female socket of the power supply. {As depicted in Figure~{\ref{expSetup}}(a), the upper part of PS \#1 (50 mm inength) is made of polyoxymethylene (POM) and theower part (45 mm) is made of glass. This design allows the visualization of the discharge inside the reactor, showing that the plasma is generated only over the quartz tube, from where it is subsequently directed to the metallic connector. PS \#2 is made entirely of POM.}

\begin{table}[htb] 
\caption{Dimensions (in mm) of the DBD reactors used with the plasma sources \#1 and \#2.\label{reactorDim}}
\centering
\begin{tabular}{p{80mm} cc}
\hline
\hline
\textbf{Dimension}	& \textbf{PS \#1}	& \textbf{PS \#2}\\
\hline
Enclosure outerength & 95.0 & 80.0 \\
Enclosure innerength ($l_i$ ) & 70.0 & 65.0 \\
Enclosure outer diameter & 28.0 & 12.0 \\
Enclosure inner diameter & 24.0 & 10.0 \\
Quartz tubeength (inside the chamber, $l_q$ ) & 62.0 & 54.0 \\
Quartz tube outer diameter & 6.0 & 4.0 \\
Quartz tube inner diameter & 4.0 & 2.0 \\
Electrode diameter & 3.3 & 1.8\\
\hline
\hline
\end{tabular}
\end{table}

All physical quantities investigated in this work were measured as a function of the $\rm{H_2}$ percentage in the working gas, which varied from 0 (without $\rm{H_2}$) to 3.5\%. For this purpose, two gas cylinders were used. One with $\rm{Ar}$ 99.999\% pure (from Air Liquide, Brazil) and another with a premixed $\rm{Ar:H_2}$ mixture in the proportion $96.5\%\rm{Ar}:3.5\%\rm{H_2}$, both 99.999\% pure (from Air Products, Brazil). The variation of the $\rm{H_2}$ percentage in the experiments was achieved by adjusting the flow of each gas coming from the cylinders with mass flow controllers (one from Horiba, model N100, and another from Omega, model FMA5520A), in a way that the total gas flow rate was kept constant and equal to 2.0 slm. The gases were then injected into a chamber to ensure proper gas mixing before entering the plasma source. Before starting the measurements, both working gases were purged for one hour to remove residual humidity from both the piping and the gas mixer.

\subsection{Electrical characterization}
The main scheme for monitoring the waveforms of applied voltage ($V(t)$) and discharge current ($i(t)$) with plasma jet impinging on a copper ($\rm{Cu}$ ) target is presented in Figure~\ref{expSetup}(a). The electrical characterizations of the device consisted mainly of obtaining the mean discharge power ($P_{dis}$) and effective discharge current ($i_{RMS}$) as a function of the $\rm{H_2}$ percentage. For this purpose, the applied voltage was measured using a 1000:1 voltage probe (Tektronix, model P6015A), and the electrical current flowing through the circuit was measured with a current monitor (Pearson, model 4100). All electrical signals were recorded with an oscilloscope (Rigol, model DS1104Z) and the $P_{dis}$ values were calculated using Eq.~\ref{eqPdis} \cite{Ashpis_progress_2017, Pipa_equivalent_2019}:
\begin{equation}
 P_{dis} = \frac{1}{\tau} \int _0^\tau V(t) i(t) dt \label{eqPdis}
\end{equation}

\noindent where $\tau$ = 1.7 ms and 16.67 ms for the plasma sources \#1 and \#2, respectively.

\subsection{Measurements of gas temperatures}
Gas temperature ($T_g$) measurements were carried out by employing a fiber optic temperature (FOT) sensor, from Weidmann Technologies Deutschland GmbH, Germany, with 0.2 K precision. The schematic for the $T_g$ measurements is depicted in Figure~\ref{expSetup}(c). The FOT sensor is connected to a transducer (from the same manufacturer), which was set to collect 25 temperature values in an interval of 0.5 s between consecutive measurements. The data acquisition for gas temperature measurements was carried out as a function of the hydrogen percentage in the gas mixture with the plasma jet on throughout data collection. The plasma jet was impinging on the $\rm{Cu}$ plate set at a distance $d =$ 10 mm from the gas/plasma outlet. In addition, when measuring $T_g$, the FOT sensor was placed at a fixed position, in a direction parallel to the surface of the $\rm{Cu}$ plate at a distance of 1.0 mm, with only the tip of the FOT sensor touching the plasma jet.

\subsection{Optical emission spectroscopy diagnostics}
Optical emission spectroscopy (OES) was employed for the identification of emitting species, as well as for the analysis of the intensity behavior of selected emitting species. From theatter, {for electron density ($n_e$) was estimated, and rotational and vibrational temperatures ($T_r$ and $T_v$, respectively) values of nitrogen molecules were calculated. All of these parameters were studied} as a function of the amount of hydrogen in the gas mixture. The details of the OES measurements are depicted in Figure~\ref{expSetup}(d). The OES measurements were carried out with the plasma jet impinging perpendicularly on a metal grid (made of commercial iron) placed over a quartz plate. Theight emitted by the plasma jet was collected with a collimatingens (20 mm in diameter), placed after the quartz plate in the axial direction, and conducted to the spectrometer through an optical fiber. A multichannel Avantes spectrometer (model AvaSpec-ULS2048X64T), with spectral resolution (FWHM) equal to 0.76 {\textpm} 0.02 nm, was employed for an overview of the emission spectra. More detailed spectroscopic measurements were performed with a Horiba multichannel spectrometer (model MicroHR), with FWHM equal to 0.34 {\textpm} 0.01 nm. {The spectroscopic measurements for calculations of $T_r$, $T_v$ and $n_e$ were carried out using the Horiba spectrometer.}

{Rotational and vibrational temperatures  of the plasma plume were obtained by the spectral emission from the $\rm{N_2}$ molecules. For this purpose, we used the second positive system $\rm{N_2}$, specifically the $C {}^{3} \Pi_u, \nu' \rightarrow B {}^{3} \Pi_g , \nu''$ transition, with $\Delta \nu$ = $\nu' - \nu''$ = -2, spanning the wavelength range from 362 to 382 nm }\cite{moon_comparative_2003,bruggeman_gas_2014,zhang_determination_2015,ono_optical_2016}.
{Spectra simulations were carried out using the massiveOES software }\cite{vorac_batch_2017,vorac_state-by-state_2017}.{ The temperature values are determined by finding the simulated spectra that provide the best fit to the measured spectra.}

{The electron density ($n_e$) of the $\rm{Ar-H_2}$ atmospheric pressure plasma jet was obtained from the Stark broadening ($w_S$) of the hydrogen Balmer-$\beta$ine ($H_\beta$, $\lambda$ = 486.13 nm). The expression that correlates the electron density to the Stark broadening of the $H_\beta$ine emission is given by }\cite{gigosos_computer_2003, faltynek_electron_2016}:
\begin{equation}
 w_S = 4.8 \left({\frac{n_e}{10^{23}}}\right)^{0.68116}~(nm) \label{eqWs}
\end{equation}

\noindent {being that Eq.{~\ref{eqWs}} is valid for $n_e$ and values ranging from $1 \times 10^{20}~m^{-3}$ to $1 \times 10^{25}~m^{-3}$ and for a wide range of electron temperature values.}

{The measured spectraline was fitted with a Voigt profile, which is a convolution of a Gaussian profile (due to Doppler effect and instrumental broadening) and a Lorentzian contribution (due to both Stark and van der Waals effects) }\cite{Hofmann_power_2011, rooij_electron_2024}.
{While the instrumental broadening ($w_I$) is constant, the Doppler broadening ($w_D$) depends on the gas temperature ($T_g$). The expression for $w_D$ as a function of $T_g$ for the $H_\beta$ine emission can be written as }\cite{faltynek_electron_2016, Hofmann_power_2011}:

\begin{equation}
 w_D = 3.4817 \times 10^{-4} (T_g)^{1/2}~(nm) \label{eqWd1}
\end{equation}

{Then, the total Gaussian to the broadenedine is given by:}

\begin{equation}
 w_G = (w_D^2 + w_I^2)^{1/2}~(nm) \label{eqWg}
\end{equation}

{The van der Waals broadening ($w_W$) is, in essence, a parameter that depends on the gas temperature and can be expressed as }\cite{faltynek_electron_2016, Hofmann_power_2011,Yubero_measuring_2015}:

\begin{equation}
 w_W = \frac{C}{T_g^{7/10}}~(nm) \label{eqWw}
\end{equation}

{where $C$ is a constant equal to 5.24 for the $H_\beta$ine emission in plasmas formed using argon as the working gas }\cite{Hofmann_power_2011}.
{Thus, the total Lorentzian broadening is:}

\begin{equation}
 w_L = w_S + w_W~(nm) \label{eqWl}
\end{equation}

{Many works write an approximate expression for the full width at half maximum (FWHM) of the Voigt profile ($w_V$). In this work, the fitting of $H_\beta$ine emission was performed using the \textit{VoigtFit} routine implemented in the \textit{Lmfit} Python package }\cite{Newville_lmfit_2016, Krogager_voigtfit_2018}{. The $w_G$ value is calculated using a previously measured $T_g$ value before the fitting, and defined as a constant in the \textit{VoigtFit} routine. In this way, $w_L$ is determined by finding the values that best fit the experimental data. Then, the $w_W$ value is calculated using the same $T_g$ value used to obtain $w_D$ and {Eq.~\ref{eqWl}} is applied to isolate $w_S$ from $w_L$.}

\subsection{Water activation}
\begin{SCfigure}[1][ht]
\centering
\includegraphics[width=0.4\textwidth]{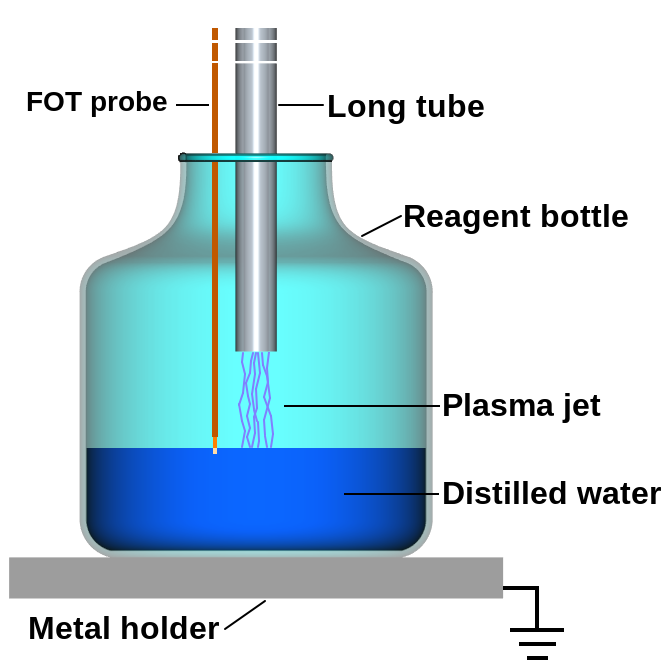}
\caption{Experimental setup used for water treatment. \label{expApplic}} 
\end{SCfigure}

Plasma jets generated using  $\rm{Ar}$ and $\rm{Ar + H_2}$ as the working gases were directed to water samples to evaluate possible differences between the two processes. The schematic for water exposure to the plasma jet is depicted in Figure~\ref{expApplic}. 30 ml samples of commercially available distilled water, placed in an open glass bottle, were exposed to plasma treatment for 5 and 10 min for PS \#1 and for 10 min for PS \#2. In all experiments, the distance between the plasma outlet and the target surface was 10.0 mm. Moreover, the plasma treatment was performed with the plasma jet impinging on a fixed position of the water surface.

Absorption spectroscopy of plasma-treated water was performed using a UV-Vis spectrometer from Perkin Elmer (model Lambda25) in the wavelength range from 190 nm to 350 nm. The RONS formed in theiquid phase were then investigated by fitting the experimental data with absorbance data of various RONS ($f_j = f_{{RONS}_j}$) as a function of the wavelength ($\lambda$). The absorbance data were extracted from works that presented absorbance as a function of the wavelength for RONS in aqueous medium \cite{Barrett_absorption_1972, Ling_uv-vis_2013, Oh_-situ_2016, Liu_quantifying_2019, oh_uvvis_2017}. The equation used to perform the curve fitting has the following general form:
\begin{equation}
 g(\lambda) = \sum _j m_j f_j(\lambda) \label{eqASfit}
\end{equation}

\noindent where $m_j$ are multipliers of $f_j(\lambda)$ and $g(\lambda)$ is the resulting curve. If the UV-Vis system used to measure the absorption spectrum has absorbance calibration, then the $m_j$ multipliers return the concentrations of each of the RONS in the plasma-treated water \cite{oh_uvvis_2017, Petkovic_assessment_2024}. The fitting procedure is done using the Levenberg-Marquardt (\textit{leastsq}) method implemented in the minimize function of the Lmfit package \cite{Newville_lmfit_2016}. {The quality of the fit for each UV-Vis spectrum was quantitatively assessed by calculating the chi‑square ($X^2$) and the coefficient of determination ($R^2$);ower values of $X^2$ with $R^2$ close to 1 indicate that the model reproduces well the measured spectrum. To verify that all assigned reactive species contributed significantly to the fit result, aeave‑one‑out cross‑validation was performed: each species was sequentially omitted from the fitting basis, then the fit was recomputed and the resulting variations in $X^2$ and $R^2$ are recorded. Therefore, if any omitted component causes an increase in the $X^2$ value by more than 20\%, even if the variation in $R^2$ is small, it is assumed that the corresponding species is required to reproduce the observed absorbance profile.} {\interfootnotelinepenalty=10000\footnote{For comparison purposes, all the species in the fitting basis are displayed in the results. However, those that do not influence significantly in the fit result are displayed without its corresponding uncertainty value}}
Semi-quantitative tests were also performed for the detection of $\rm{H_2O_2}$, $\rm{NO_2^{-}}$ and $\rm{NO_3^{-}}$ in the plasma treated water (PTW) using test strips (from Merck, Canada, for $\rm{H_2O_2}$ and from Quantofix, Germany, for the other two). {The test strips were also used for an initial cross-check of the fitting method. It was done by comparing the value of the area under the curves of $\rm{H_2O_2}$, $\rm{NO_2^{-}}$ and $\rm{NO_3^{-}}$ - obtained by fitting the UV-Vis spectra - to the values of the test strip readings after plasma exposure.} Besides that, for the detection of ammonia ($\rm{NH_3}$) in the PTW, a semi-quantitative analysis based on a freshwater test kit (from Alcon, Brazil) was employed.
The temperature right below the water surface was also monitored as a function of exposure time. For this purpose, the FOT sensor was placed parallel to the plasma jet axis, 10 mm away, with its tip immersed in water to a depth of approximately 1 mm, as depicted in Figure~\ref{expApplic}. The water temperature measurements were performed when the plasma jet was on, as well as with only the gas flow impinging on the water surface. performed when the plasma jet was on, as well as with only the gas flow impinging on the water surface.

\section{Results and discussion}
Most of the results presented in this section were obtained using only the following two conditions: pure Ar ($\rm{Ar}$ (100\%)) as the  working gas and the admixture $\rm{Ar} (96.5\%):\rm{H_2}(3.5\%)$. Thus, for practical purposes, from now on the first will beabeled simply as $\rm{Ar}$ while theast will beabeled as $\rm{Ar + H_2}$. It will be indicated in the text when different $\rm{Ar-H_2}$ proportions are used. An important information about the measurements as a function of the $\rm{H_2}$ percentage is that, except when differently indicated, the measurements started with $\rm{H_2}$ = 0 \%, and the plasma sources were kept in continuous operation until each set of measurements was finished.

\subsection{Electrical characterization of the discharge}
\begin{figure}[htb]
\centering
\begin{minipage}{0.4\textwidth}
\centering
\includegraphics[width=\textwidth]{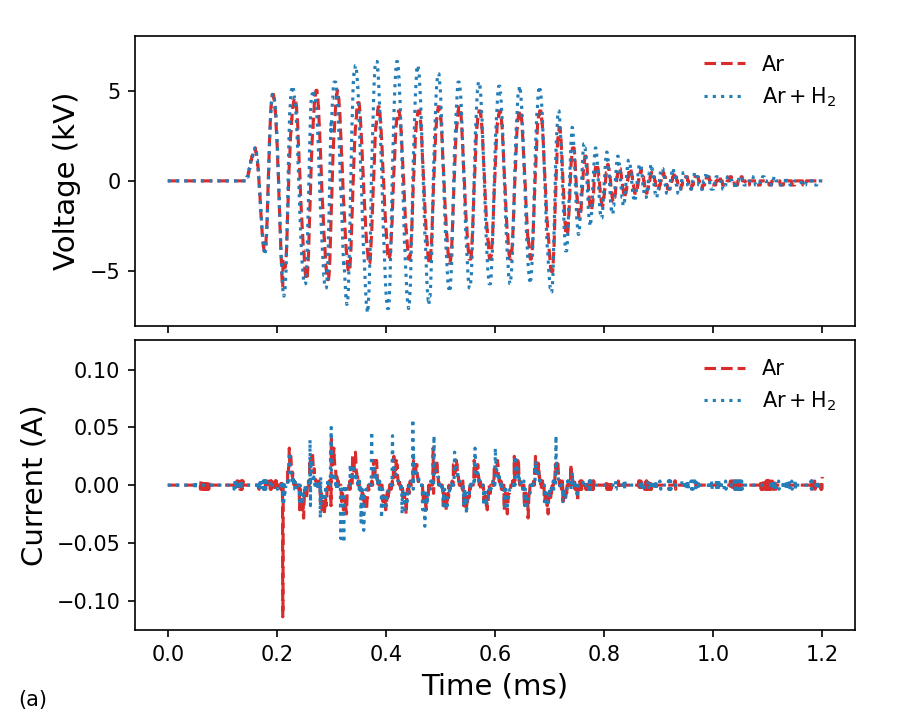}
\end{minipage}
\begin{minipage}{0.4\textwidth}
\centering
\includegraphics[width=\textwidth]{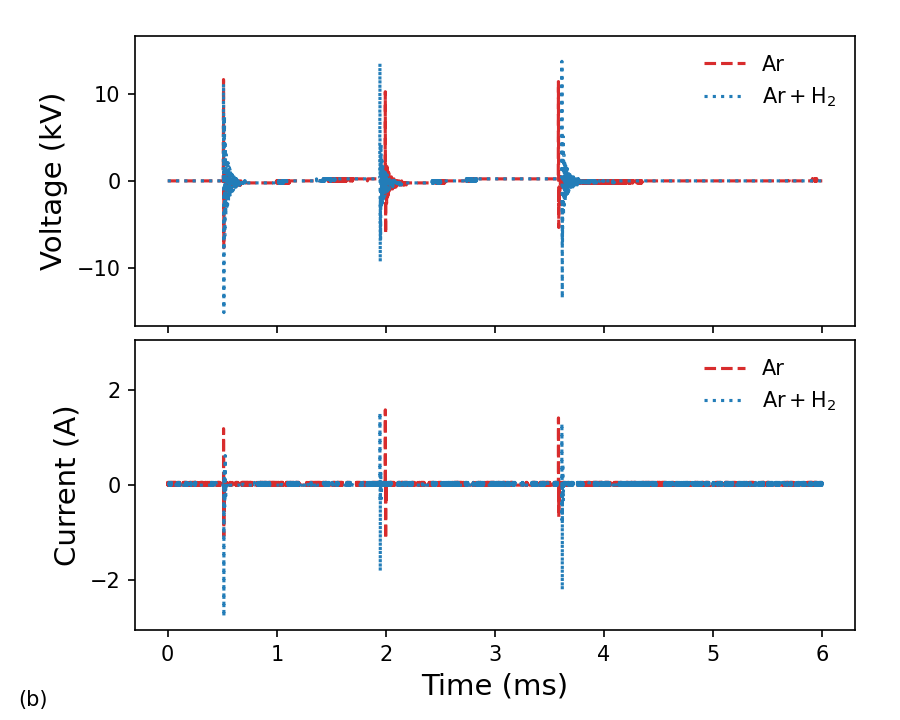}
\end{minipage}
\caption{Voltage ($V$) and current ($i$) waveforms measured using $\rm{Ar}$ and $\rm{Ar + H_2}$ for (a) PS \#1 and (b) for PS \#2, respectively. \label{typwf}}
\end{figure}

In Figure~\ref{typwf} are shown the characteristic voltage ($V(t)$) and current ($i(t)$) waveforms for both $\rm{Ar}$ and $\rm{Ar + H_2}$ working gases. In Figure~\ref{typwf} (a) it is shown the data for the plasma source \#1 and in Figure~\ref{typwf} (b) for \#2. It can be seen that for both plasma sources, the amplitudes of the voltage and the current signals were clearly affected by the presence of $\rm{H_2}$ in the working gas. Variations in voltage waveforms are usually associated with a change in the system impedance. In the present case, the impedance seems to be affected by the different discharge behavior observed in the presence or not of $\rm{H_2}$ in the working gas.

\begin{SCfigure}[1][ht]
\centering
\includegraphics[width=0.5\textwidth]{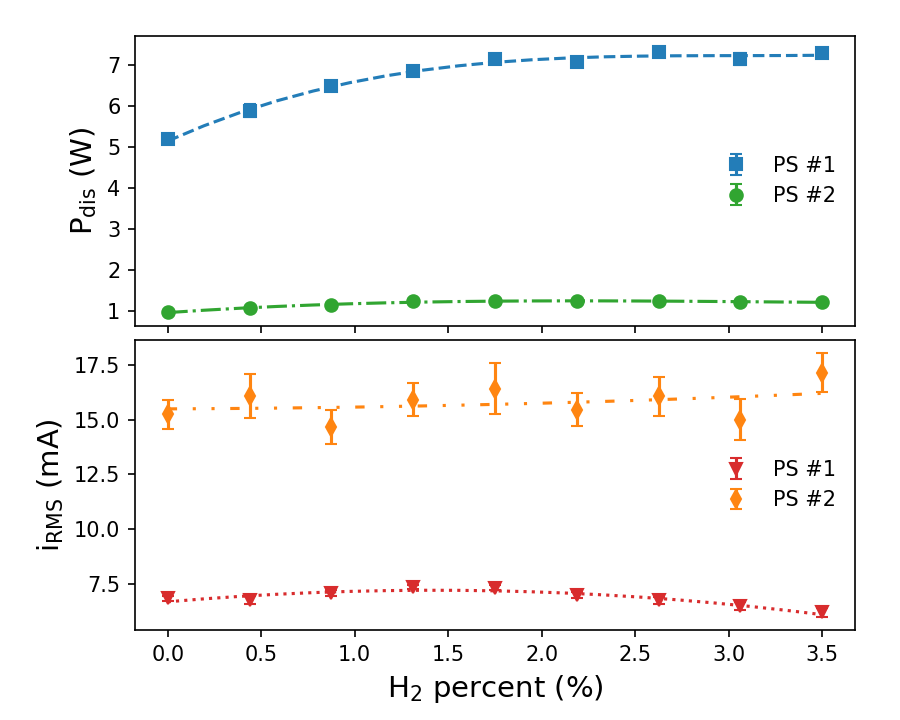}
\caption{Discharge power and effective current as a function of the hydrogen content in the working gas for plasma sources (a) and \#2. \label{PiVSperc}}
\end{SCfigure}

In Figure~\ref{PiVSperc} are presented the calculated values of $P_{dis}$ and $i_{RMS}$ as a function of the $\rm{H_2}$ content in the working gas for (a) PS \#1 and (b) PS \#2. As can be seen, the PS \#1 delivers a significantly higher power, which makes it appropriate for material processing only. On the other hand, PS \#2 exhibits discharge power slightly above 1.0 W, so it is also suitable for biomedical treatments. Regarding the discharge power, the behavior of the $P_{dis}$ curves as a function of the $\rm{H_2}$ percent is similar, increasing when the $\rm{H_2}$ content increases from 0 to $\sim$1.5\%, with a saturation trend after that. The behavior of the $i_{RMS}$ curves as a function of the $\rm{H_2}$ fraction slightly differs for the two plasma sources. Unlike the $P_{dis}$ values, the PS \#2 exhibits higher \textit{rms} current values, which are almost constant within the measurements' uncertainties for this source. On the other hand, when PS \#1 is used, the $i_{RMS}$ curve peaks when the $\rm{H_2}$ content is nearly 1.5\% ,  and  beyond this point, it tends to decrease.

\subsection{OES and thermal diagnostics}
\subsubsection{Spectra overview and emission intensity for selected species}
\begin{figure}[htb]
\centering
\begin{minipage}{0.53\textwidth}
\centering
\includegraphics[width=\textwidth]{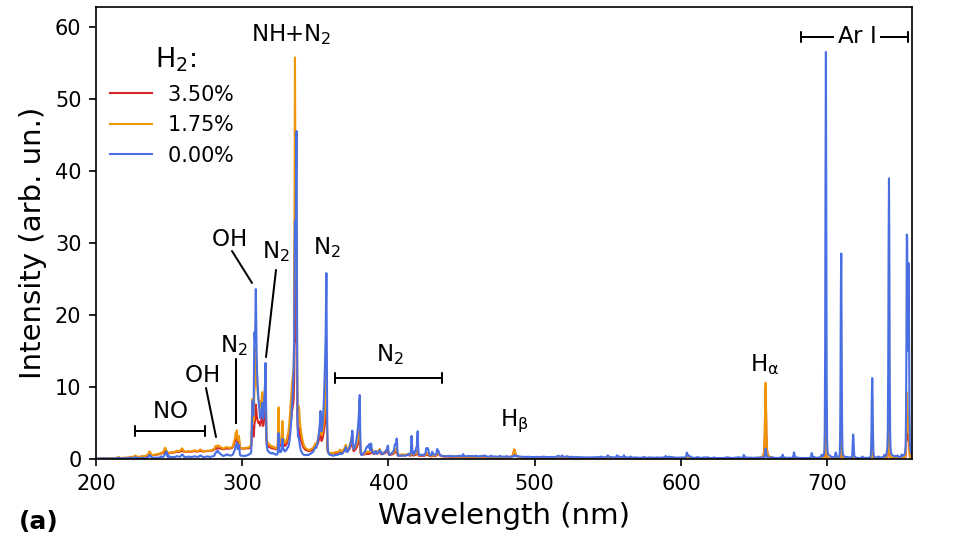}
\end{minipage}
\begin{minipage}{0.46\textwidth}
\centering
\includegraphics[width=\textwidth]{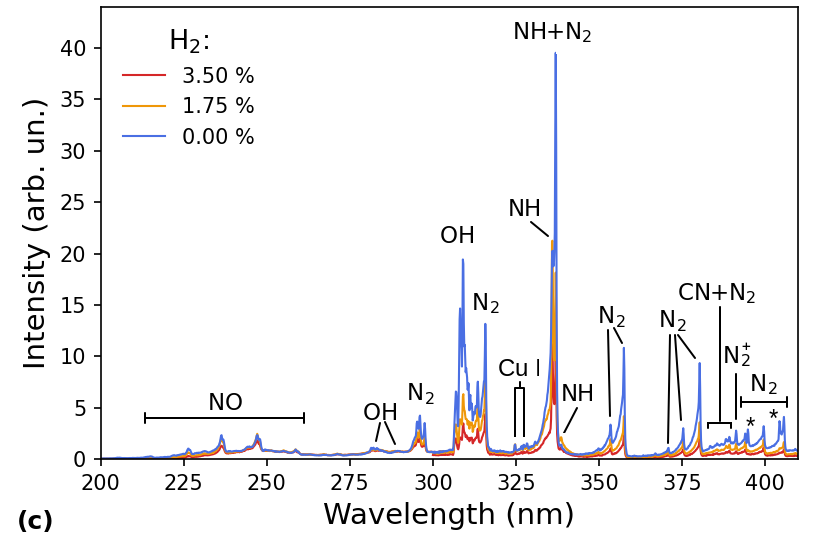}
\end{minipage}
\\
\begin{minipage}{0.53\textwidth}
\centering
\includegraphics[width=\textwidth]{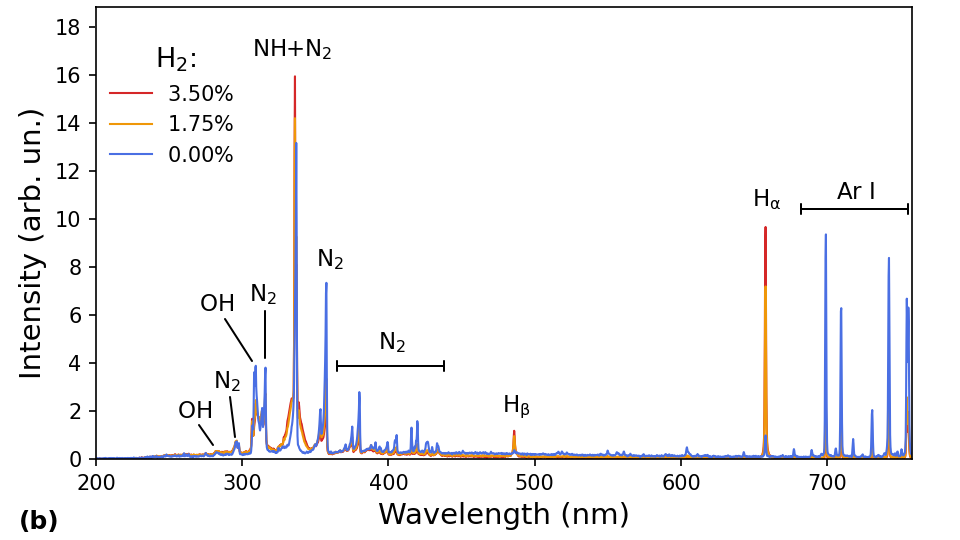}
\end{minipage}
\begin{minipage}{0.46\textwidth}
\centering
\includegraphics[width=\textwidth]{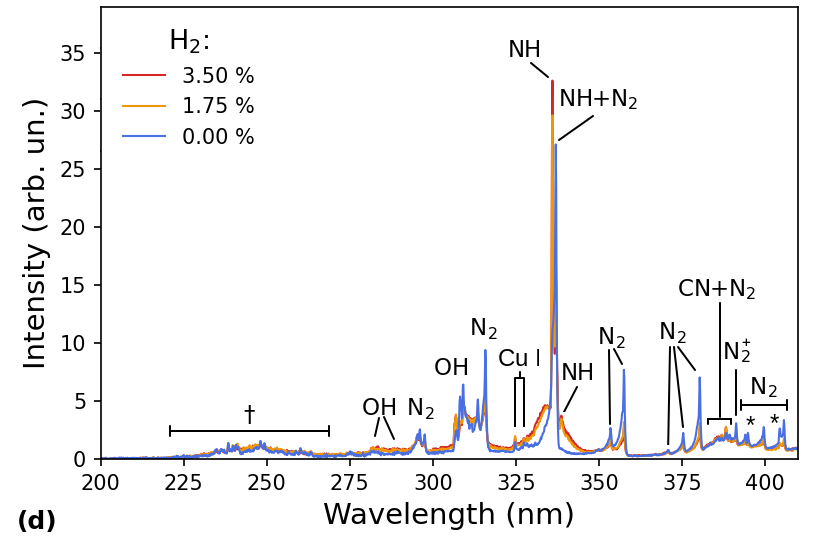}
\end{minipage}
\caption{{Overview of the optical emission spectra in the wavelength range from 200 nm to 758 nm measured for plasma sources (a) \#1 and (b) \#2. Detailed spectra from 200 nm to 410 nm were obtained for plasma sources \#1 and \#2, (c) and (d), respectively. Curves for three different $\rm{H_2}$ percentages are shown in each figure. The spectra in (a) and (b) were measured with the Avantes spectrometer, while the spectra in (c) and (d) were measured using the Horiba spectrometer. Each spectrum shown in (c) and (d) is composed of two sets of measurements made between 200-300 nm and 300-410 nm. The asterisks in (c) and (d) denote $\rm{Ar~I}$ine emissions. The $\dagger$ in (d) indicates an overlap of multiple emitting species.} \label{oesOvervw}}
\end{figure}

Wide-range spectra of the optical emission coming from the plasma jet powered by the plasma sources \#1 and \#2 are shown in Figure~\ref{oesOvervw}, (a) and (b), respectively. More detailed spectra in the wavelength range from 200 nm to 410 nm are presented in Figure~\ref{oesOvervw} (c) and (d), for plasma sources \#1 and \#2, respectively. Each subfigure shows the spectra obtained for three different hydrogen concentrations in the gas mixture (0\%, 1.75\%, and 3.5\%). From Figure~\ref{oesOvervw} it can be seen that the main molecular emissions come from $\rm{NO}$, $\rm{OH}$, $\rm{N_2}$, $\rm{NH}$ and $\rm{CN}$, as well as the molecular nitrogen ions ($\rm{N_2^{+}}$), while atomic emissions come from $\rm{H}$, $\rm{Ar}$ and $\rm{Cu}$. The $\rm{Cu}$ emissions come from $\rm{Cu}$ atoms detached from the copper wire due to the plasma-wire interaction. Atomic emissions from $\rm{O}$ (777 nm and 844 nm triplet systems) were also observed using the Horiba spectrometer, but the spectra of that region, which contains mostly $\rm{Ar~I}$ emissions, are not shown here for the sake of clarity. Band emissions from the $\rm{NO}$ radical are {clearly} observed only for plasma source \#1. {For plasma source \#2, the intensity of $\rm{NO}$ emissions is too weak and the bands are overlapped by multiple emitting species, which probably come from iron oxide and $\rm{Fe^{0+/1+}}$ species. Theower $\rm{NO}$ emission observed for PS \#2 is an expected result, since the $\rm{NO}$ production in APPJs depends on the discharge power}\cite{Pipa_absolute_2008}. All other emitting species are clearly present in the spectra measured with both plasma sources. It is interesting to note that traces of $\rm{NH}$ emissions as well as the atomic Hydrogenines are observed even if no $\rm{H_2}$ is added to the working gas. In this case, the $\rm{NH}$ radicals areikely formed from the H atoms dissociated from water molecules present in theaboratory environment, whose relative humidity is typically of the order of 60\%.

The $\rm{NH}$ band emissions observed in the emission spectra come from the $A {}^3 \Pi_i \rightarrow X {}^3 \Sigma_{-}$ system. The most intense $\rm{NH}$ emissions are from the Q-branch at 336.0 nm ($\nu' = 0, \nu'' = 0$) and 337.0 nm ($\nu' = 1, \nu'' = 1$) \cite{pearse_identification_1950,lents_evaluation_1973}. As it can be seen in Figure~\ref{oesOvervw}, the $\rm{NH}$ emissions overlap the emissions from the $\rm{N_2}$ $C {}^3 \Pi_u \rightarrow B {}^3 \Pi_g$ system at 337.1 nm ($\nu' = 0, \nu'' = 0$). From Figure~\ref{oesOvervw} (d) it can be noticed that the R- and P-branches of the $\rm{NH}$ $A {}^3 \Pi_i \rightarrow X {}^3 \Sigma_{-}$ system (ateft and right of the Q-branch, respectively) are probably being excited as well. The emission peaking at 338.7 nm also comes from the Q-branch of the $\rm{NH}$ $A {}^3 \Pi_i \rightarrow X {}^3 \Sigma_{-}$ system ($\nu' = 2, \nu'' = 2$) \cite{Plemmons_analysis_1998}.

It is important to mention that the integration time used in the OES measurements was changed according to the plasma source and the spectrometer used. Thus, a direct comparison among the intensity values measured for different sets can not be done. The integration time was adjusted in a way that the emission peaks/bands have enough definition when averaged over 50 measurements. Of course, at that given experimental setup, once the integration time was defined, it was kept unchanged for all OES measurements as a function of the $\rm{H_2}$ fraction. Therefore, a comparison between theight intensity obtained for different H2 content can be made within this specific setup. 

\begin{figure}[htb]
\centering
\begin{minipage}{0.4\textwidth}
\centering
\includegraphics[width=\textwidth]{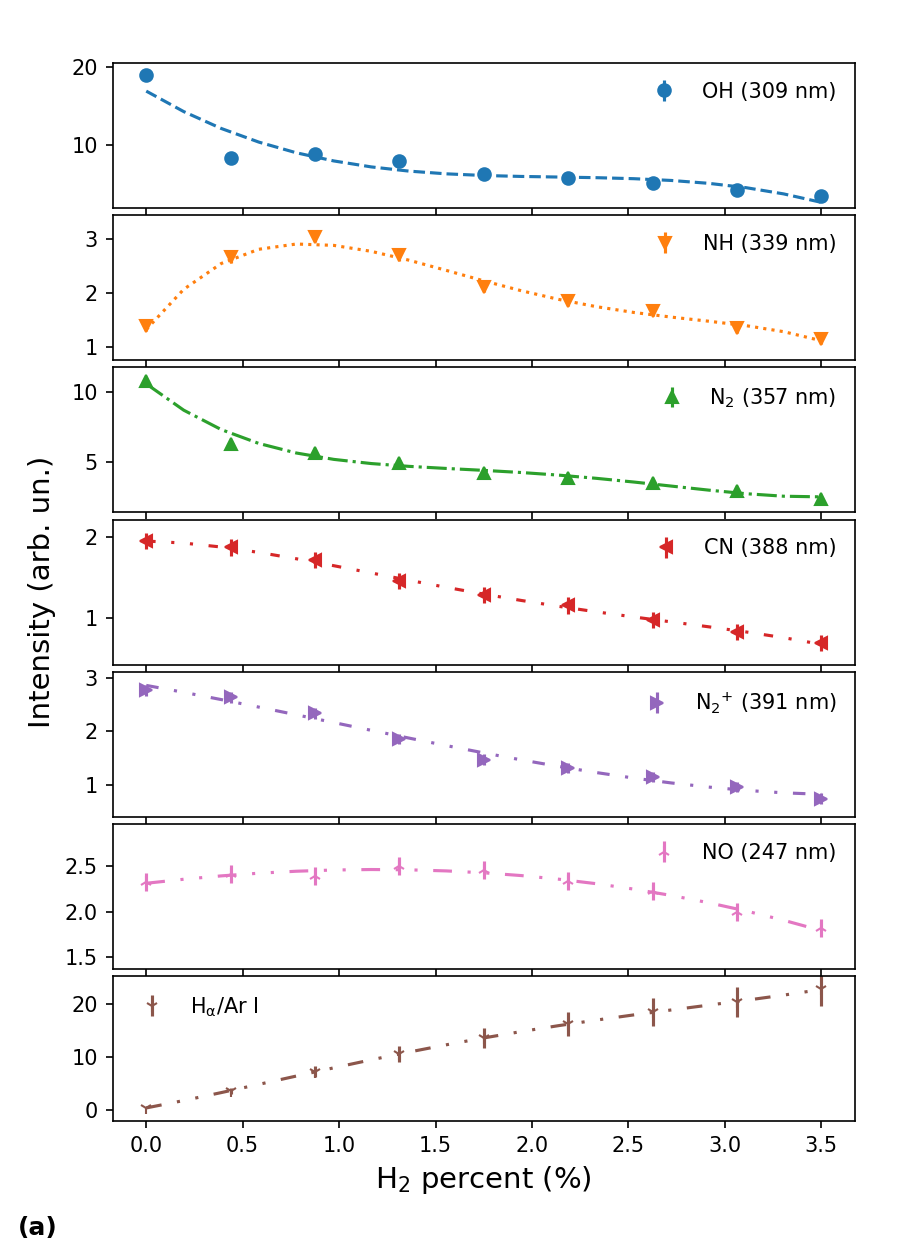}
\end{minipage}
\begin{minipage}{0.4\textwidth}
\centering
\includegraphics[width=\textwidth]{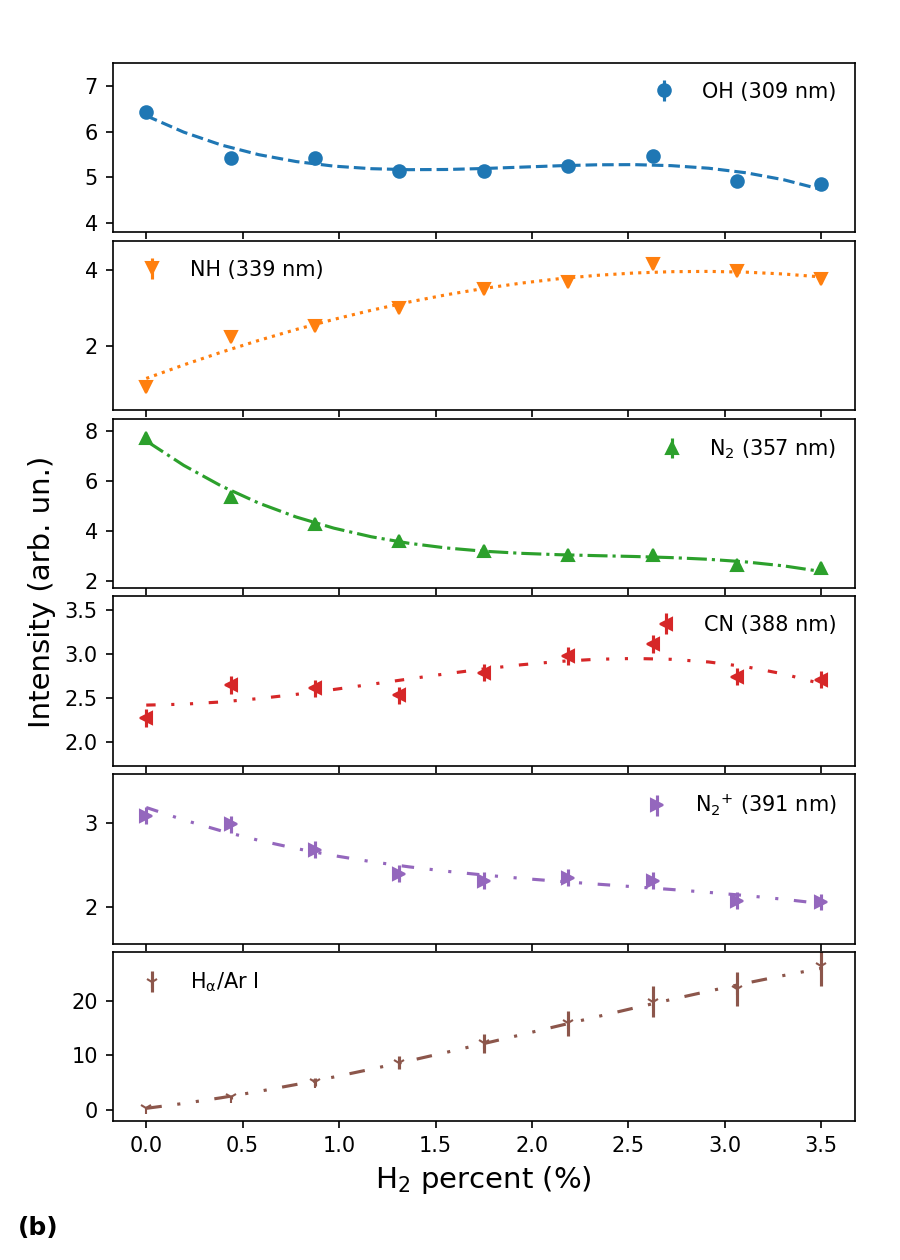}
\end{minipage}
\caption{Radiation intensity of selected species as a function of the hydrogen content in the gas mixture for (a) PS \#1 and (b) PS \#2. The ratio between the intensity of $\rm{H_{\alpha}}$ and $\rm{Ar~I}$ (727 nm)ine emissions is also plotted in theowest frame of (a) and (b). OES measurements were performed with the Horiba spectrometer.\label{OESIntensities}}
\end{figure}

Figure~\ref{OESIntensities} presents curves of radiation intensity as a function of the $\rm{H_2}$ percent in the working gas for selected molecular band emissions. From that figure, it can be seen that the emission of some excited molecular species as a function of the $\rm{H_2}$ content in the gas mixture behaves differently depending on the plasma source. Regarding the first one, it is interesting to note that the intensity curve for $\rm{NH}$ in Figure~\ref{OESIntensities} (a) presents a peak value at $\sim$0.9\% of $\rm{H_2}$ in the gas mixture. On the other hand, the intensity curves for $\rm{OH}$, $\rm{N_2}$, $\rm{CN}$ and $\rm{N_2^{+}}$ decrease monotonically as more $\rm{H_2}$ is added to the mixture.

From Figure~\ref{OESIntensities} (b), it can be seen that when plasma source \#2 is used, theight emission from the excited $\rm{NH}$ and $\rm{CN}$ species tends to increase as a function of the hydrogen content in the gas mixture. However, the increase in the $\rm{CN}$ight emission isess pronounced than that observed for $\rm{NH}$. On the other hand, theight emitted from the excited $\rm{OH}$, $\rm{N_2}$ and $\rm{N_2^{+}}$ exhibits a decreasing trend as more $\rm{H_2}$ is added to the gas.
Considering that the $\rm{NH}$ight emission peaks at $\rm{H_2}$$\sim$2.6\%, the best operating condition for $\rm{NH}$ production probably occurs between 2.5\% and 3.5\% $\rm{H_2}$. Nevertheless, further studies extending the $\rm{H_2}$ percent range beyond 3.5\% are required to confirm this trend and, consequently, the optimal operating point for $\rm{NH}$ production with plasma source \#2.

In theowest frames of Figure~\ref{OESIntensities} (a) and (b) are also plotted the ratio between the intensity of the $\rm{H_{\alpha}}$ and the $\rm{Ar~I}$ine emissions ($\rm{H_{\alpha}}/\rm{Ar~I}$). Interestingly, the curves obtained for both plasma sources have not only similar behavior, but also the values of the ratio are considerably close.

\subsubsection{Gas and OES temperature measurements}
\begin{SCfigure}[1][ht]
\centering
\includegraphics[width=0.5\textwidth]{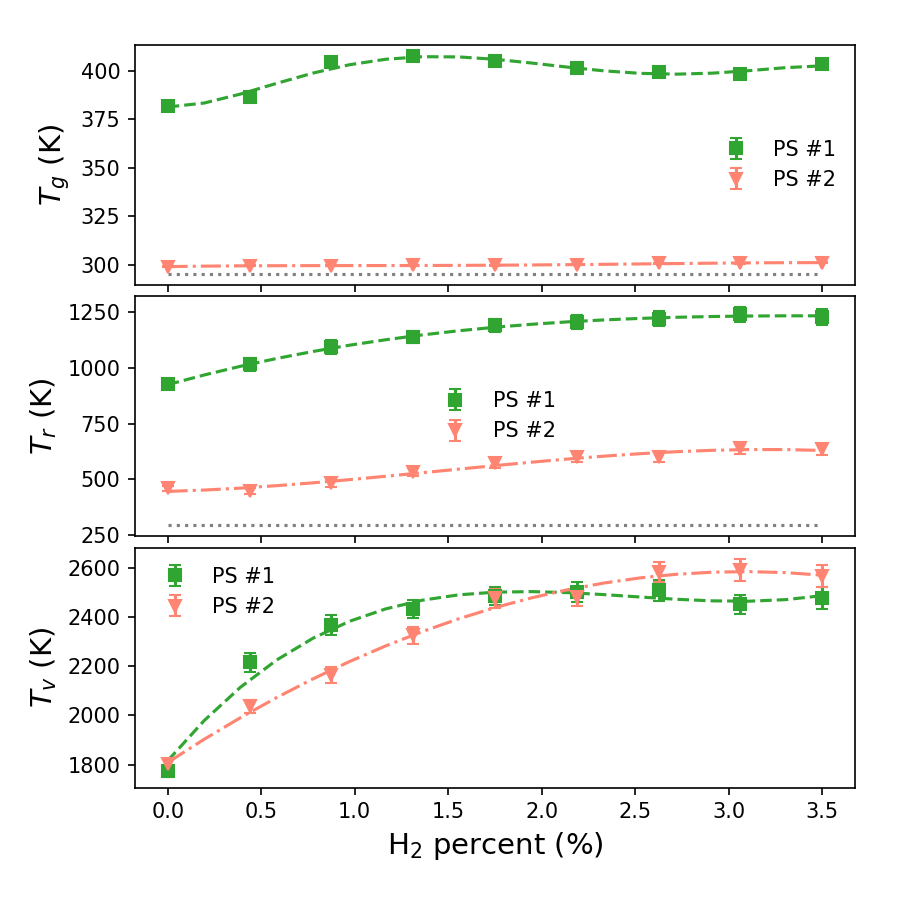}
\caption{{Variation of the different temperatures as a function of the $\rm{H_2}$ content in the gas mixture for plasma sources \#1 and \#2. Gas temperature (top), rotational temperature (middle),  and vibrational temperature(bottom) s. The room temperature is indicated by a dottedine with both $T_g$ and $T_r$ curves}. \label{TempsvsH2}}
\end{SCfigure}

In Figure~\ref{TempsvsH2} are presented the results for $T_g${, $T_r$ and $T_v$} as a function of the hydrogen content in the working gas for both plasma sources. From that figure, it can be observed that {both $T_g$ and $T_r$} values measured for plasma source \#1 are considerably higher than those obtained for plasma source \#2{, with the $T_g$ values nearly 100 K higher and $T_r$ almost the double  for PS \#1}. Regarding the behavior of $T_g${ and $T_r$} as a function of the $\rm{H_2}$ content, in all cases this parameter increases as more $\rm{H_2}$ is added to the gas. However, the variations are more pronounced for PS \#1.

Regarding the $T_g$ values as a function of the $\rm{H_2}$ content in the gas mixture, when PS \#1 is used, a small addition of $\rm{H_2}$ ($\sim$0.44\%) causes a nearly 5 K increase in $T_g$. However, increasing the $\rm{H_2}$ content to $\sim$0.88\% causes an increase of almost 20 K in $T_g$. After that, the $T_g$ values tend to reach a plateau with small oscillations around 400 K. The $T_g$ values measured for PS \#2 do not increase significantly as more $\rm{H_2}$ is added, remaining very close to the room temperature.

{An interesting result is that the $T_v$ behavior and its absolute values as a function of the $\rm{H_2}$ percent are very similar for both plasma sources, which indicates that the vibrational excitation depends more on the gas content than on the plasma source in the experiments carried out in the present work.}

\subsection{Electron density measurements}
\begin{SCfigure}[1][ht]
\centering
\includegraphics[width=0.5\textwidth]{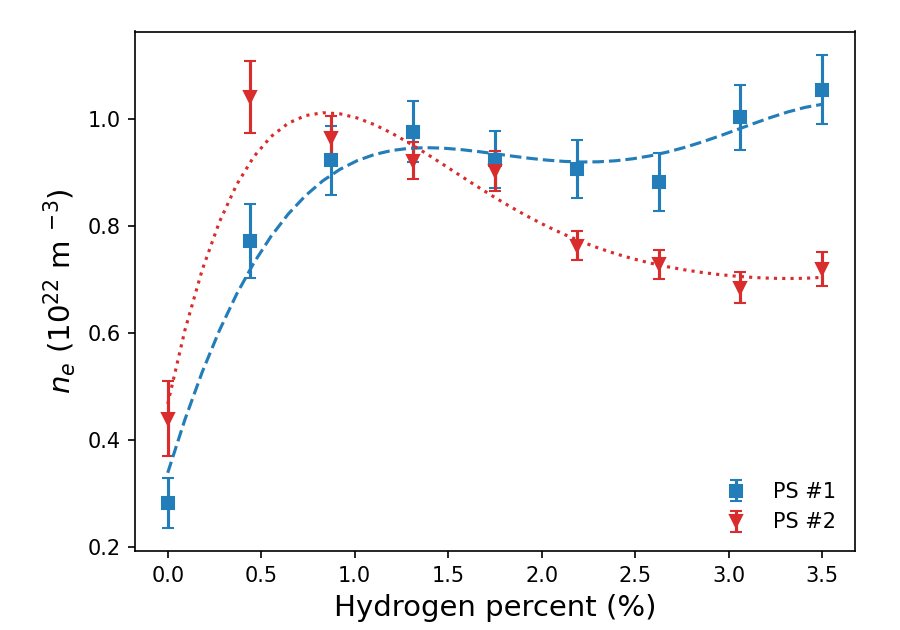}
\caption{{Variation of the electron density as a function of the $\rm{H_2}$ percentage in the plasma jet. Dashed and dottedines are trend curves.} \label{neH2}}
\end{SCfigure}

{The variation in the values of the electron density ($n_e$) as a function of the $\rm{H_2}$ percent in the gas composition is shown in Figure{~\ref{neH2}}. From that figure, the most remarkable observation to be made is that the values of $n_e$ increase considerably with the addition of only a small fraction of $\rm{H_2}$ to the working gas. It is also notable that both plasma sources reach similar $n_e$ values, of the order of $10^{21}-10^{22}~m^{-3}$, even though their discharge powers, gas temperatures, and waveforms differ considerably. This is an indication that the electron density depends more on the gas chemistry than on other parametersike the $P_{dis}$, $T_g$ and voltage waveform. The behavior of the $n_e$ curves in Figure{~\ref{neH2}} presents almost the same increase rate for a small addition of $\rm{H_2}$ to the working gas. However, while the $n_e$ values reach a plateau for PS \#1 as more $\rm{H_2}$ is added, a reduction trend is observed for PS \#2. }

{The electron density values obtained in the current work are consistent with previous results reported in theiterature for APPJs as well as in experiments employing other APP sources with $\rm{Ar}$ and/or $\rm{Ar-H_2}$ as the working gas }\cite{Guo_characteristics_2018, das_optical_2018, rooij_electron_2024, sarani_characterization_2014, srikar_development_2024}.

\subsection{Water treatment results}
\begin{figure}[htb]
\centering
\begin{minipage}{0.49\textwidth}
\centering
\includegraphics[width=0.9\textwidth]{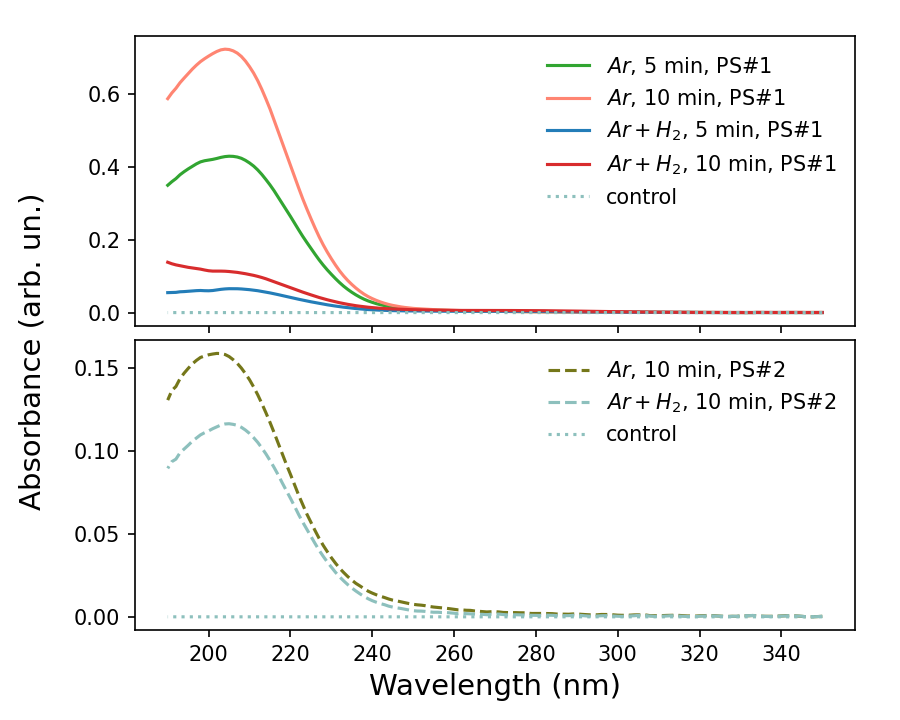}
\caption{Overview of the UV-Vis absorption spectra of PTW samples using $\rm{Ar}$ and $\rm{Ar + H_2}$ with PS \#1 for 5 and 10 min plasma exposure, and with PS \#2 for 10 min. The plasma-treated water volume was 30 ml in all cases. \label{waterTr}}
\end{minipage}
\hfill
\begin{minipage}{0.49\textwidth}
\centering
\includegraphics[width=0.9\textwidth]{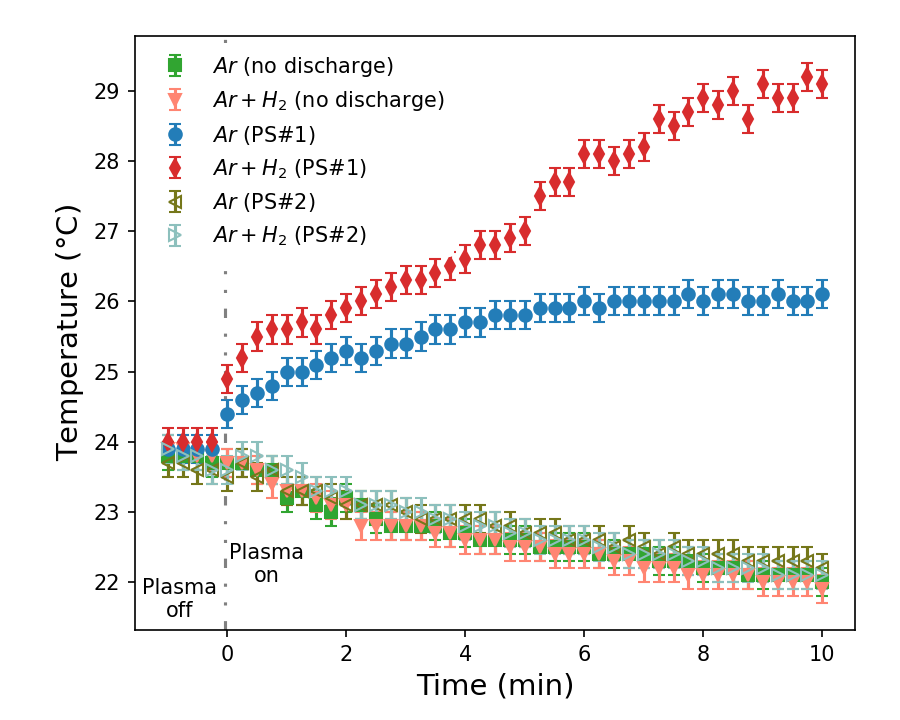}
\caption{Temporal evolution of the water temperature for $\rm{Ar}$ and $\rm{Ar + H_2}$, without plasma discharge and with plasma discharge for both plasma sources. \label{waterTempTime}}
\end{minipage}
\end{figure}

\begin{figure}[ht]
\centering
\begin{minipage}{0.35\textwidth}
\centering
\includegraphics[width=\textwidth]{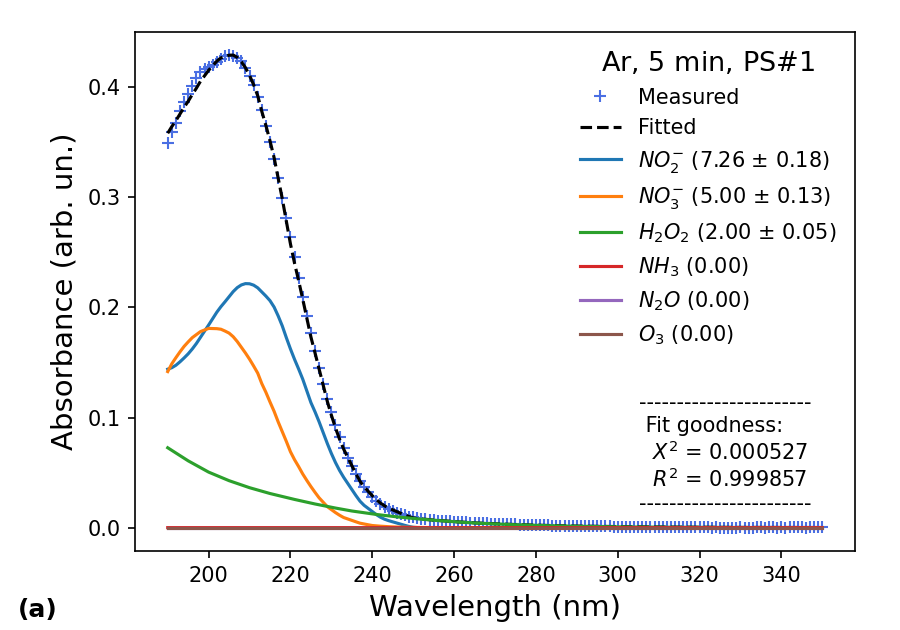}
\end{minipage}
\begin{minipage}{0.35\textwidth}
\centering
\includegraphics[width=\textwidth]{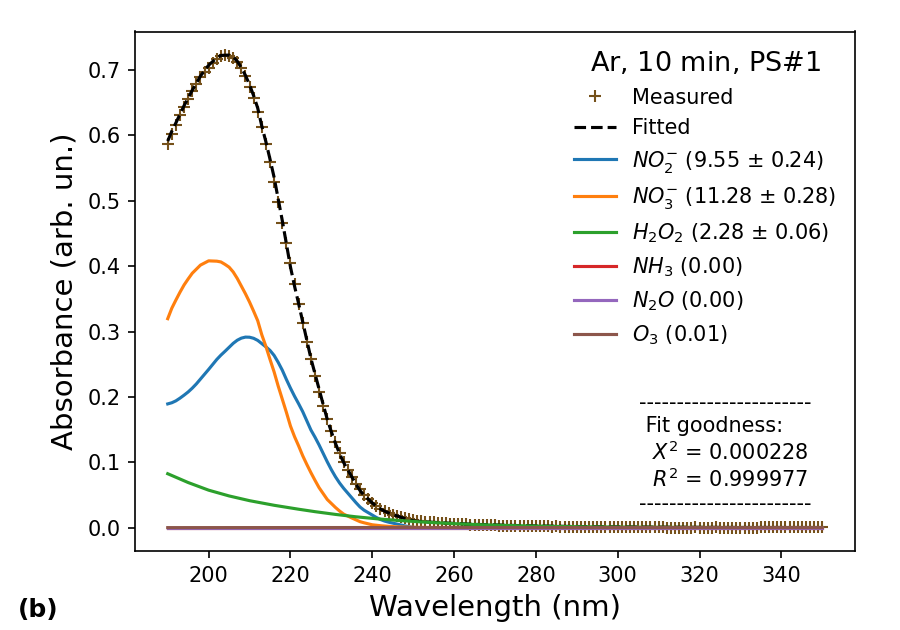}
\end{minipage}
\\
\begin{minipage}{0.35\textwidth}
\centering
\includegraphics[width=\textwidth]{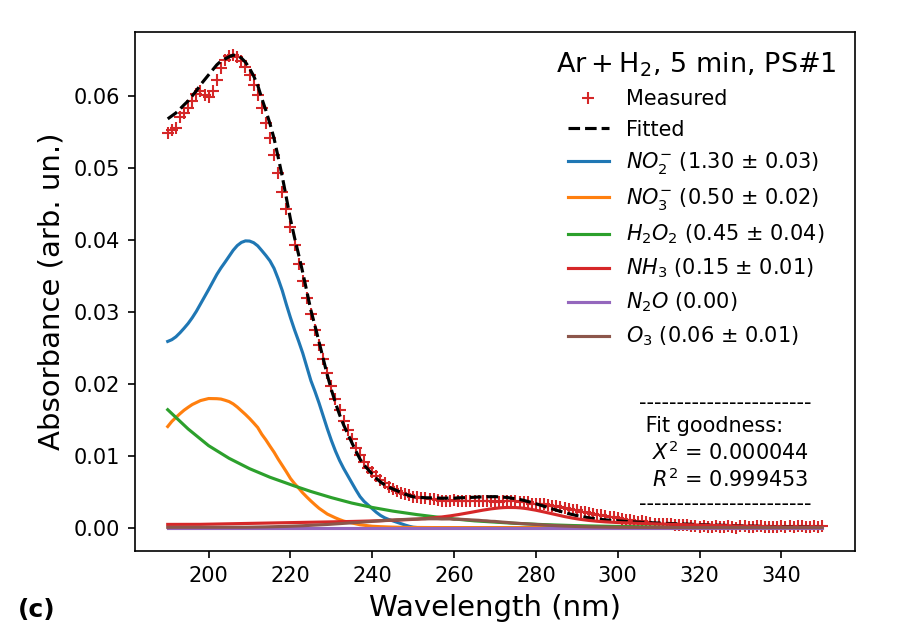}
\end{minipage}
\begin{minipage}{0.35\textwidth}
\centering
\includegraphics[width=\textwidth]{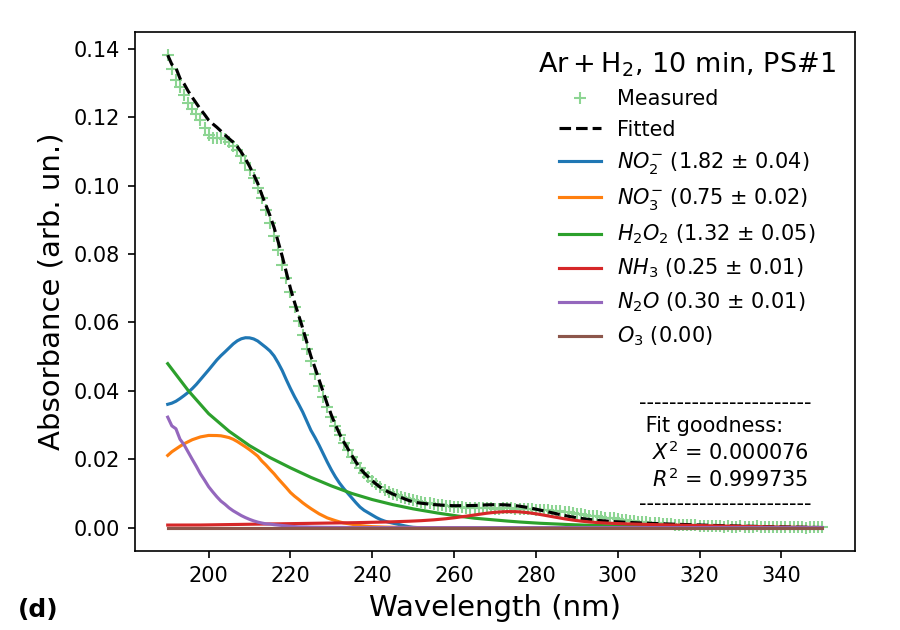}
\end{minipage}
\\
\begin{minipage}{0.35\textwidth}
\centering
\includegraphics[width=\textwidth]{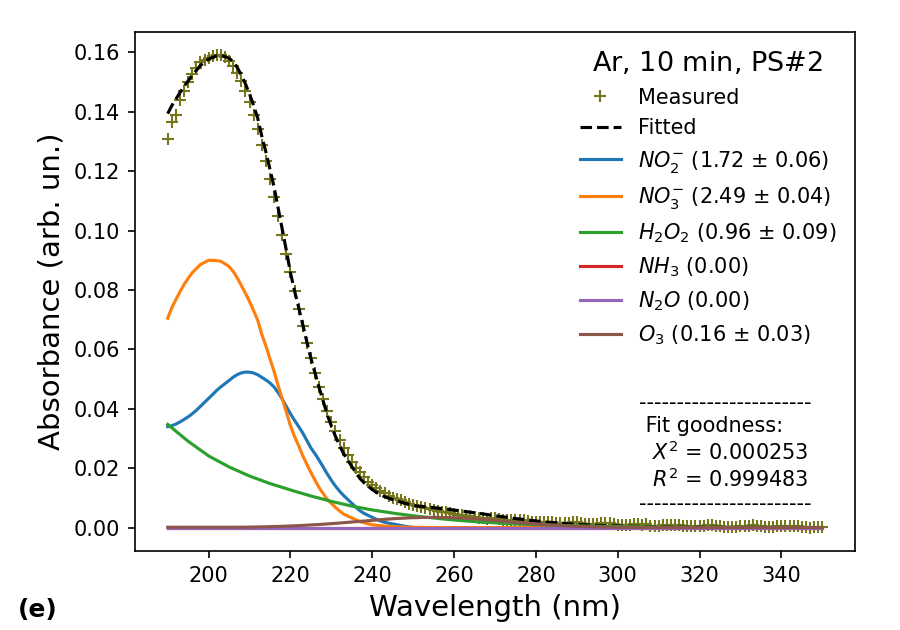}
\end{minipage}
\begin{minipage}{0.35\textwidth}
\centering
\includegraphics[width=\textwidth]{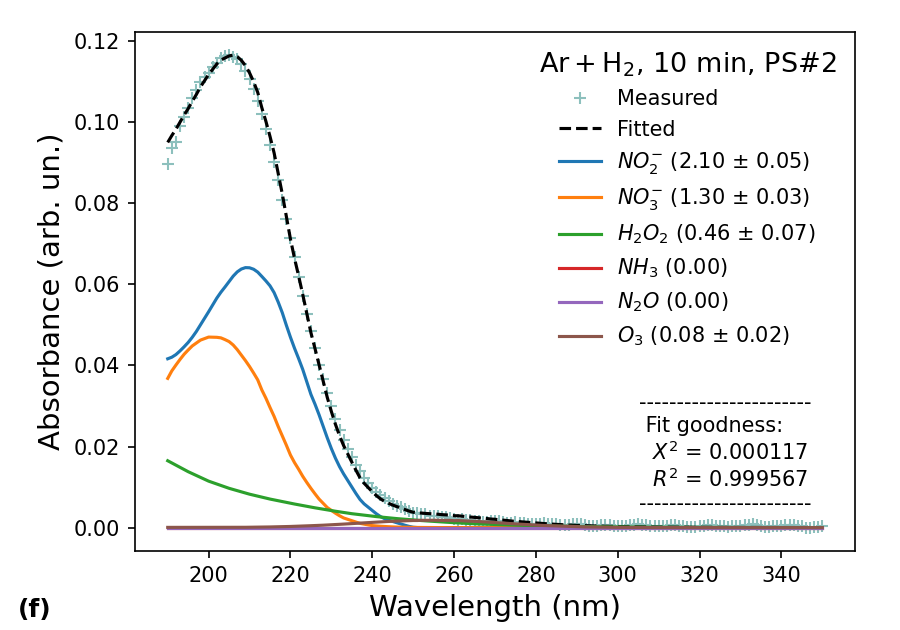}
\end{minipage}
\caption{Fits of UV-Vis spectra profiles of plasma-treated water. Using PS {\#}1: (a) Ar plasma jet for 5 minutes; (b) Ar plasma jet for 10 minutes; (c) $\rm{Ar + H_2}$ plasma jet for 5 minutes; (d) $\rm{Ar + H_2}$ plasma jet $\rm{Ar + H_2}$ for 10 minutes. Using PS {\#}2: (e) Ar plasma jet for 10 minutes; (f) $\rm{Ar + H_2}$ plasma jet for 10 minutes. The numbers in the parentheses are the values of the areas under the corresponding curves. \label{uvvisFits}}
\end{figure}

In this section, we present the results obtained when water samples were exposed to plasma. Plasma jets generated with $\rm{Ar}$ and $\rm{Ar + H_2}$ as the working gas were applied to the surface of 30 ml water samples for exposure times of 5 and 10 minutes for PS \#1 and for 10 minutes for PS \#2. An overview of the UV-Vis absorption spectra for all the samples is shown in Figure~\ref{waterTr}. The temporal evolution of the water temperature of the samples under plasma treatment is shown in Figure~\ref{waterTempTime}. Detailed analysis of the UV-Vis spectra for identification of the RONS formed in the exposed water samples is presented in Figure~\ref{uvvisFits}(a-f).
In addition, when using PS \#1, test strips were employed for semi-quantitative analysis of $\rm{H_2O_2}$, $\rm{NO_2^{-}}$ and $\rm{NO_3^{-}}$ and a freshwater test kit was used for detection of $\rm{NH_3}$. The photos of the semi-quantitative tests are shown in \ref{appendix:A} (Figure~\ref{tsPhotos} for $\rm{H_2O_2}$, $\rm{NO_2^{-}}$ and $\rm{NO_3^{-}}$ and Figure~\ref{nh3Photos} for $\rm{NH_3}$).

Figure~\ref{waterTr} clearly indicates that, in comparison to the sample that was not exposed to plasma (control),  the absorption spectra of all water samples changed after plasma treatment. From Figure~\ref{waterTr} it can be seen that when using PS \#1, the addition of $\rm{H_2}$ to the working gas changes the absorbance of the water samples, not only in intensity, but the shape of the curve is altered as well. This indicates not only a change in the amount of generated reactive species but also a change in the type of reactive species present in the treated water.

It is worth mentioning that, in addition to the data shown in Figure~\ref{waterTr}, experiments were performed with water samples exposed to the $\rm{Ar}$ and $\rm{Ar + H_2}$ gas flow without discharge ignition (no plasma formation). In those cases, no significant changes were observed in the UV-Vis spectra in comparison with the control sample. Furthermore, in preliminary tests, UV-Vis measurements at higher wavelengths ($\lambda >$ 350 nm) were also performed, and no significant changes occurred.

The curves of water temperature as a function of time presented in Figure~\ref{waterTempTime} reveal that the gas flow only, i.e., without discharge ignition, has a cooling effect on the water samples, which is probably due toiquid evaporation. This effect is, of course, the same for both working gases. On the other hand, when the plasma is ignited using PS \#1, the temperature of the water samples starts to rise. When $\rm{Ar}$ is used as the working gas, the temperature reaches a saturation value after nearly 6 minutes of plasma exposure. However, when the $\rm{Ar + H_2}$ plasma jet is employed, the temperature rise is almostinear up to 8 minutes of plasma exposure, with an indication that a plateau tends to be established after that. The difference in the water temperatures with $\rm{Ar}$ and $\rm{Ar + H_2}$ obtained for PS \#1 is a direct consequence of the gas temperature combined with the discharge power measured for each working gas, that are higher for $\rm{Ar + H_2}$ in both cases. When PS \#2 is employed in the treatment, for both working gases, the water temperature presents a slight rise at the beginning of plasma exposure. However, even with plasma on, the temperature of the water samples keeps decreasing, as in the case without a discharge. This finding indicates that theow-power source PS \#2 generates a plasma jet, whose temperature is not high enough to warm up the water sample.

Figure~\ref{uvvisFits} shows the curve fit results of the UV-Vis spectra measured for plasma jets produced with (a-d) PS \#1 and (e-f) PS \#2. By using the fitting procedure described in section 2.5 it was possible to identify the main RONS diluted in the PTW samples as well as to separate the absorbance curves for each case. The numbers in the parentheses in Figure~\ref{uvvisFits} are the values of the areas under the corresponding RONS absorbance curves, which areinked to the total absorbance of the species.

The values of the area under the curves of the $\rm{H_2O_2}$, $\rm{NO_2^{-}}$ and $\rm{NO_3^{-}}$ in Figures~\ref{uvvisFits} (a) and (b) were used not to directly quantify the amount of such species, but as a complementary tool to be compared to the results obtained with the test strips. For example, by comparing the values of the areas under the $\rm{H_2O_2}$ curves in Figures~\ref{uvvisFits} (a) and (b), $\rm{Ar}$ - 5 min and $\rm{Ar}$ - 10 min, with the respective test strip results in Figure~\ref{tsPhotos} it can be verified that the measured values agree in terms of magnitude. The same comparison can be carried out for $\rm{NO_2^{-}}$ and $\rm{NO_3^{-}}$ species: the results from the areas under the curves of each species agree in terms of magnitude with the results from the test strips. With this, the process of identifying the species formed in the water after exposing the samples to plasma becomes more reliable.

With that said, the analysis of the plasma-treated water samples reveals that when PS \#1 is employed and $\rm{Ar + H_2}$ is the working gas, not only $\rm{H_2O_2}$, $\rm{NO_2^{-}}$ and $\rm{NO_3^{-}}$ are formed in the water, but also $\rm{O_3}$ and $\rm{NH_3}$ for 5 minutes of plasma exposure. When the plasma exposure time is 10 minutes, no more $\rm{O_3}$ was found, but nitrous oxide ($\rm{N_2O}$) is detected, and the amount of $\rm{NH_3}$ is increased.

Since some chemical processes can occur at higher temperatures, the formation of different species in the water samples for different plasma exposure times observed for the $\rm{Ar + H_2}$ gas mixture (see Figure~\ref{uvvisFits} (c and d)) is probablyinked to the variation in the water temperature presented in Figure~\ref{waterTempTime}. This dependence on water temperature for $\rm{NH_3}$ formation is partially confirmed by the results obtained with PS \#2. In this case, the water temperature is muchower than with PS \#1, and no $\rm{NH_3}$ formation occurred when the $\rm{Ar + H_2}$ gas mixture was used, even after exposing the water to plasma for 10 minutes. 

{Regarding total absorbance for PS \#1 with and without $\rm{H_2}$ in the gas mixture (compare Figures{~\ref{uvvisFits}}(a) with{~\ref{uvvisFits}}(c) and {~\ref{uvvisFits}}(b) with{~\ref{uvvisFits}}(d)), the differences are consistent with the reduced concentrations of $\rm{H_2O_2}$, $\rm{NO_2^{-}}$, and $\rm{NO_3^{-}}$ found in the presence of $\rm{H_2}$. The reduction of such species in the case of $\rm{Ar+H_2}$ plasma isikely due to their consumption in ammonia formation. 

Comparing the results in Figures{~\ref{uvvisFits}}(b) and{~\ref{uvvisFits}}(e), it can be seen that the total absorbance obtained for PS \#1 is higher than that for PS \#2. The primary factor for this result is the effective pulse repetition rate of each plasma source: $\approx$588 Hz for PS \#1 and 180 Hz for PS \#2. This parameter directly determines the effective plasma treatment duration, which is higher for PS \#1, resulting in the formation ofarger quantities of RONS species. Additional contributing factors for the better performance of PS \#1 with Ar as the working gas include its higher discharge power, gas temperature, and rotational temperature, as well as its significant $\rm{NO}$ production, which is negligible for PS \#2. These parameters act synergistically to enhance the water treatment efficacy using PS \#1.

For the results obtained using PS \#2 (Figures{~\ref{uvvisFits}}(e) and{~\ref{uvvisFits}}(f)), the near-identical absorbance with and without $\rm{H_2}$ addition is consistent with the minimal changes $P_{dis}$ and $T_g$, as well as with the fact that the plasma generated by PS \#2 does not meet the conditions required for ammonia production in either the gas or aqueous phase.}

The production of $\rm{NH_3}$ with PS \#1 may have occurred during the aqueous or gaseous phases. 
{In the gas phase, one of the main mechanismseading to ammonia formation is the direct hydrogenation of nitrogen atoms, starting with ionization/dissociation of $\rm{N_2}$ and $\rm{H_2}$, from the ambient air and from the working gas, respectively, followed by stepwise hydrogenation }\cite{van_duc_long_understanding_2022}:

\begin{equation}
  \begin{split}
    e + N_2 \rightarrow 2N + e \\
    e + H_2 \rightarrow 2H + e \label{eqDiss}
  \end{split}
\end{equation}
\begin{equation}
  \begin{split}
    N + H \rightarrow NH \\
    NH + H \rightarrow NH_2 \\
    NH_2 + H \rightarrow NH_3 \label{swhydro}
  \end{split}
\end{equation}

{Then the $\rm{NH_3}$ formed in the gas phase dissolves in water ($\rm{NH_3~_{(aq)}}$). This reaction pathway is partially supported by the appearance of $\rm{NH}$ radicals in the OES spectra in the presence of $\rm{H_2}$ in the working gas. In addition, this would explain the behavior of the $\rm{NH}$ emission intensity curve as a function of the $\rm{H_2}$ content presented in Figure{~\ref{OESIntensities}} (a), which reaches its peak for an amount of $\rm{H_2}$ close to 0.5\%, but is strongly reduced when increasing the $\rm{H_2}$ proportion in the gas mixture.}

{In the aqueous phase, due to the presence of $\rm{NO_2^{-}}$ and $\rm{NO_3^{-}}$, as well as the UVight emission from the plasma jet, two possible mechanismseading to the production of ammonia or ammonium ($\rm{NH_4^{+}}$) are }\cite{Schullehner_nitrate_2017, Yang_formation_2021}:

\begin{equation}
    NO_3^{-} + 8e^{-}_{(aq)} + 10H^{+} \rightarrow NH_4^{+} + 3H_2O \label{NO3NH4}
\end{equation}
\noindent and
\begin{equation}
    2NO_2^{-} + 2H_2O + 4H^{+} \rightarrow 2NH_4^{+} + 3O_2 \label{NO2NH4}
\end{equation}

{Thus, $\rm{NH_4^{+}}$ can react with $\rm{OH^{-}}$ to form $\rm{NH_3}$:}
\begin{equation}
    NH_4^{+} + OH^{-} \rightarrow NH_3 + H_2O \label{NH4OH}
\end{equation}

{This reaction pathway is supported by the UV-Vis spectra presented in Figure{~\ref{uvvisFits}}(a-d). When comparing the results obtained for the same treatment time, the amounts of $\rm{NO_2^{-}}$, $\rm{NO_3^{-}}$, and $\rm{H_2O_2}$ are much higher in the samples treated only with $\rm{Ar}$ as the working gas. The reduction in the amounts of $\rm{NO_2^{-}}$ and $\rm{NO_3^{-}}$ that occurs when the plasma jet contains $\rm{H_2}$ is an indication that these species were consumed in reactions  {\ref{NO3NH4}} and {\ref{NO2NH4}}. And the reduction in the amount of $\rm{H_2O_2}$ may indicate that the $\rm{OH^{-}}$ radicals were consumed via reaction {\ref{NH4OH}}.}

{Considering reactions {\ref{eqDiss}} to {\ref{NO2NH4}}, two of the parameters observed in the plasma jet that appear to be particularly important for $\rm{NH_3}$ production are the amount of $\rm{NH}$ and the electron density. From Figure{~\ref{neH2}} it can be seen that the electron density obtained for PS \#2 is nearly 20\%ower than the value obtained for PS \#1 at 3.5\% of $\rm{H_2}$ in the gas composition. In this condition, the effects of dissociation/ionization of $\rm{H_2}$/$\rm{N_2}$ molecules followed by stepwise hydrogenation areess pronounced for PS \#2 than for PS \#1. Based on the $T_g$ and $T_r$ values obtained for each of the plasma sources (see Figure{~\ref{TempsvsH2}}, both higher for PS \#1 than for PS \#2, we can conclude that these two parameters, at relatively high values, are also important for $\rm{NH_3}$ production. Therefore, when using PS \#2, it is possible that the conditions for $\rm{NH_3}$ production in the gas phase are not being reached at a sufficient rate for $\rm{NH_3}$ molecules to reach the water. In addition, it is also possible that reactions  {\ref{NO3NH4}} and {\ref{NO2NH4}} are dependent on the water temperature, which would reinforce their role in $\rm{NH_3}$ production.}
{Furthermore, several reaction pathwaysead to the formation of $\rm{NO_2^{-}}$ and $\rm{NO_3^{-}}$ in the aqueous phase. In any case, the presence of $\rm{NO}$ in the gas phase has a significant influence on the amount of $\rm{NO_2^{-}}$ and $\rm{NO_3^{-}}$ detected in the water.}\cite{Bae_nitric_2024, Ghorui_atmospheric_2025} {This is another factor that explains the absence of $\rm{NH_3}$ in water samples treated with the $\rm{Ar + H_2}$ plasma jet using PS \#2, since the $\rm{NO}$ production with PS \#2 is not as significant as with PS \#1.}

Due to the presence of $\rm{NH_3}$ in the PTW when PS \#1 was employed, the use of $\rm{Ar + H_2}$ to generate plasma jets may not be recommended for medical applications in such working conditions, since $\rm{NH_3}$ is a toxic compound for some human tissues. However, it may be a good alternative for agricultural applications, since $\rm{NH_3}$ is one of the main compounds used in fertilizers.

{Both the semi‑quantitative analysis (Figure{~\ref{nh3Photos}}) and the UV‑Vis fitting (Figure{~\ref{uvvisFits}}(c)) indicate that ammonia appears only after a 5‑minute exposure of the water to the $\rm{Ar + H_2}$ plasma jet; after 10 minutes, the $\rm{NH_3}$ production is increased, and $\rm{N_2O}$ also becomes detectable. The plasma source PS \#1, which operates at a higher frequency, reaches an average discharge power of $\approx$7 W when 3.5\% $\rm{H_2}$ is added (Figure{~\ref{PiVSperc}}). For the 5‑minute treatment of 30 mL of water, this corresponds to an energy input of roughly 2.1 kJ, or $\approx$70 kJ/L. From the results obtained with the  freshwater test kit, the concentration of $\rm{NH_3}$ after 5 minutes of plasma exposure is approximately 0.5 mg/L. This gives an energy yield for $\rm{NH_3}$ production of $\approx$7 {\textmu}g/kJ, which is farower than industrial processes synthesis but potentially useful forocalized, on‑site fertilizer applications where transport of bulk chemicals is impractical. On the other hand, even in the ammonia‑producing regime, the PTW still presents $\rm{NO_2^{-}}$, $\rm{NO_3^{-}}$,  $\rm{H_2O_2}$ and, eventually, $\rm{O_3}$, which can act as mild plant‑growth stimulants or antimicrobial agents. Therefore, this combination of RONS and $\rm{NH_3}$ (also possibly $\rm{NH_4^{+}}$) could work synergistically for provision and biostimulatory activity rather than serving solely as a conventional fertilizer.}
{Regarding the production of $\rm{N_2O}$, it is an environmental drawback but not a fatal flaw of the plasma treatment. It highlights the complexity of the plasma chemistry and the need to optimize the process in order to maximize the $\rm{NH_3}$ production and avoid $\rm{N_2O}$ formation.}

\section{Conclusions}

This work deals with the electrical, thermal, and optical characterization of two plasma jets operating with $\rm{Ar + H_2}$ gas mixtures and their application in the production of plasma-activated water. Discharge and plasma properties of two plasma sources, which differ in dimensions and power supply, were analyzed as a function of the $\rm{H_2}$ content. The plasma source operating at a higher frequency (PS \#1) presentedarger values for discharge power and gas temperature when compared to the device operated at aower frequency (PS \#2).

The functionalization of water samples by plasma jet presented significant differences when employing $\rm{Ar}$ or $\rm{Ar + H_2}$ as a working gas when the water treatment was performed with PS \#1. In that case, UV-Vis spectroscopy and test strips revealed that, in comparison with the $\rm{Ar + H_2}$ plasma jet, the one operated with pure $\rm{Ar}$ed to the formation of higher concentrations of $\rm{NO_2^{-}}$, $\rm{NO_3^{-}}$ and $\rm{H_2O_2}$ in the treated samples. On the other hand, the use of $\rm{Ar + H_2}$ plasmaed to the formation of different reactive species in the PTW - namely $\rm{NH_3}$, and $\rm{N_2O}$ for 10 minutes of plasma exposure. The formation of $\rm{NH_3}$ in the PTW can be explored for agricultural purposes. When the water activation was performed with PS \#2, there were differences in the treatment results obtained with $\rm{Ar}$ or $\rm{Ar + H_2}$. However, the differences were restricted only to the amount of each RONS detected in the water samples (no different species were observed).

This work has demonstrated that atmospheric pressure plasma jets operating with a $\rm{Ar + H_2}$ mixture exhibit interesting features that can be useful in medicine and agriculture. {The role of $\rm{NH}$ radicals in plasma medicine needs further investigation  so that the application of the $\rm{Ar + H_2}$ gas mixture can be used for this purpose.}
{By using PS \#1, only a modest concentration of ammonia was generated in the PTW – possibly enough for seed germination or hydroponic supplementation, but not to replace bulk fertilizers at present. However, since $\rm{NH_3}$ production was a consequence of this study and not its primary objective, there is considerable room to improve, optimize, and scale the process.}
This can be partially accomplished by adjusting operating parameters that were kept fixed in this work,ike distance-to-target, gas flow rate,{ voltage amplitude and frequency of PS \#1}, as well as increasing the portion of $\rm{H_2}$ in the gas mixture.

\vspace{6pt} 

\section*{Data Availability Statement}
The data are contained in this manuscript. Raw data are available from the authors upon reasonable request.

\section*{Conflict of Interest}
The authors declare that there is no conflict of interest in this work.

\section*{Acknowledgments}
This work was supported by the São Paulo Research Foundation - FAPESP (grants \#2019/05856-7 and \#2020/09481-5) and by the Coordenação de Aperfeiçoamento de Pessoal de Nível Superior - CAPES. KGK acknowledges financial support from the National Council for Scientific and Technological Development - CNPq, under grant \#310608/2021-0.

\section*{References}
\bibliographystyle{unsrt}
\bibliography{water_activation}

\section*{Appendix}
\appendix
\section{Photos of the semi-quantitative analysis of PTW\label{appendix:A}}
In this appendix are presented the photos showing the results of the semi-quantitative analysis of PTW using test strips, for $\rm{H_2O_2}$, $\rm{NO_2^{-}}$ and $\rm{NO_3^{-}}$, and a freshwater test kit, for $\rm{NH_3}$. The results presented here are for PS \#1 only.

In Figure~\ref{tsPhotos} are displayed the test strip results for $\rm{H_2O_2}$, $\rm{NO_2^{-}}$ and $\rm{NO_3^{-}}$ detection. The results of $\rm{NH_3}$ measurements using the freshwater test kit are shown in Figure~\ref{nh3Photos}. Figure~\ref{nh3Scale} shows the scale for $\rm{NH_3}$ measurements. In Figure~\ref{nh3Comparison}  examples of incorrect and correct $\rm{NH_3}$ readings are displayed. Note that theighting in the photos is not as good as theighting observed with the naked eye, with theatter being much better. Therefore, the results presented in Figure~\ref{nh3Photos} refer to readings taken with the naked eye.

\setcounter{figure}{0} \renewcommand{\thefigure}{A.\arabic{figure}} 

\begin{figure}[htb]
\centering
\begin{minipage}{0.475\textwidth}
\includegraphics[width=\textwidth]{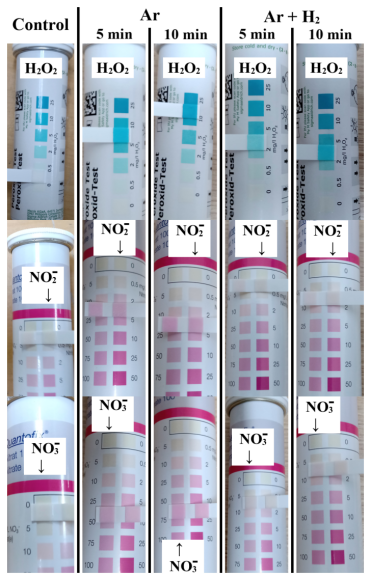}
\caption{Photos showing the results for $\rm{H_2O_2}$, $\rm{NO_2^{-}}$ and $\rm{NO_3^{-}}$ detection with test strips. \label{tsPhotos}}
\end{minipage}
\hfill
\begin{minipage}{0.505\textwidth}
\includegraphics[width=\textwidth]{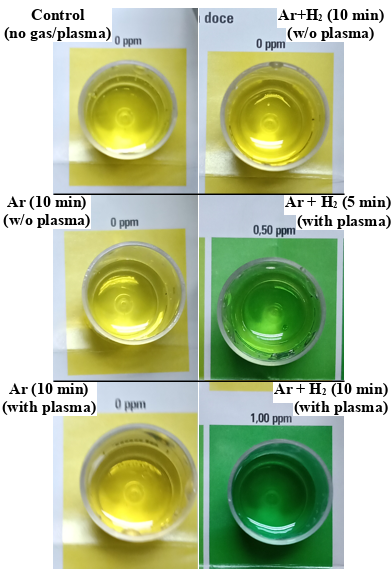}
\caption{Photos of $\rm{NH_3}$ detection with the freshwater test kit. \label{nh3Photos}}
\end{minipage}
\end{figure}

\begin{figure}[htb]
\centering
\begin{minipage}{0.46\textwidth}
\includegraphics[width=\textwidth]{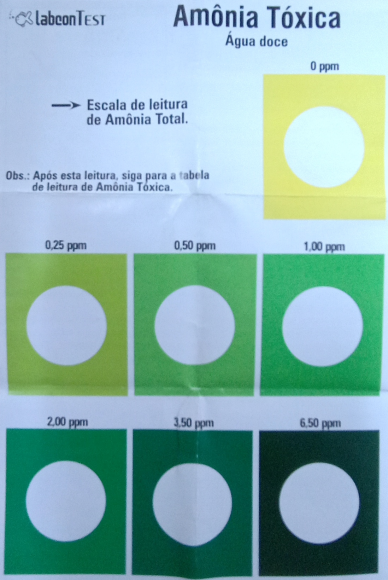}
\caption{Reference scale for $\rm{NH_3}$ detection with the freshwater test kit. \label{nh3Scale}}
\end{minipage}
\hfill
\begin{minipage}{0.42\textwidth}
\includegraphics[width=\textwidth]{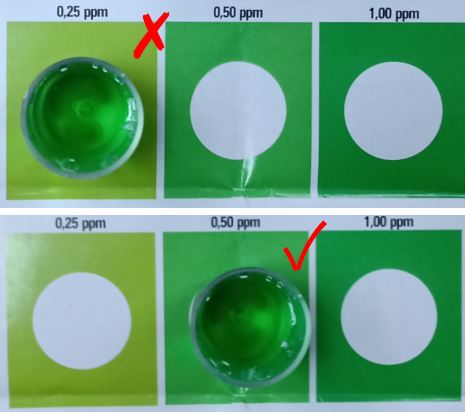}
\caption{Example of correct and incorrect $\rm{NH_3}$ estimation. Theighting in the photos does not correspond to theighting observed with the naked eye (theatter is always better). \label{nh3Comparison}}
\end{minipage}
\end{figure}

\end{document}